\documentclass[smallextended]{svjour3arxiv}

\usepackage{graphicx}
\usepackage{dcolumn}
\usepackage{bm}
\usepackage{hyperref}
\usepackage[mathlines]{lineno}
\usepackage{blkarray}
\usepackage{array}
\usepackage{bbold}
\usepackage{gensymb}
\usepackage{multirow}
\usepackage{hhline}
\usepackage{color}
\usepackage{epsfig}
\usepackage{mathptmx}
\usepackage{amsmath}
\usepackage{amssymb}
\usepackage[numbers,sort&compress]{natbib}
\newcolumntype{C}[1]{>{\centering\let\newline\\\arraybackslash\hspace{0pt}}m{#1}}
\newcommand{\kket}[1]{|{#1}\rangle}

\begin{document}


\title{Hyperfine spin qubits in irradiated malonic acid: heat-bath algorithmic cooling}

\author{Daniel K. Park \and Guanru Feng \and Robabeh Rahimi \and St\'{e}phane Labruy\`{e}re \and Taiki Shibata \and Shigeaki Nakazawa \and Kazunobu Sato \and Takeji Takui \and Raymond Laflamme \and Jonathan Baugh}

\institute{D.K. Park \and G. Feng \and R. Rahimi \and S. Labruy\`{e}re \and R. Laflamme \and J. Baugh \at Institute for Quantum Computing, University of Waterloo, Waterloo, Ontario, N2L 3G1, Canada
\and
D.K. Park \and G. Feng \and R. Rahimi \and R. Laflamme \and J. Baugh \at Department of Physics and Astronomy, University of Waterloo, Waterloo, Ontario, N2L 3G1, Canada
\and
T. Shibata \and S. Nakazawa \and K. Sato \and T. Takui \at Department of Chemistry and Molecular Materials Science, Graduate School of Science, Osaka City University, Sumiyoshi-ku, Osaka, 558-8585, Japan
\and
R. Laflamme \at Perimeter Institute for Theoretical Physics, Waterloo, Ontario, N2J 2W9, Canada
\at Canadian Institute for Advanced Research, Toronto, Ontario M5G 1Z8, Canada
\and
J. Baugh \at Department of Chemistry, University of Waterloo, Waterloo, Ontario, N2L 3G1, Canada \\\email{baugh@uwaterloo.ca}
}

\maketitle

\begin{abstract}
The ability to perform quantum error correction is a significant hurdle for scalable quantum information processing. A key requirement for multiple-round quantum error correction is the ability to dynamically extract entropy from ancilla qubits. Heat-bath algorithmic cooling is a method that uses quantum logic operations to move entropy from one subsystem to another, and permits cooling of a spin qubit below the closed system (Shannon) bound. Gamma-irradiated, $^{13}$C-labeled malonic acid provides up to 5 spin qubits: 1 spin-half electron and 4 spin-half nuclei. The nuclei are strongly hyperfine coupled to the electron and can be controlled either by exploiting the anisotropic part of the hyperfine interaction or by using pulsed electron-nuclear double resonance (ENDOR) techniques. The electron connects the nuclei to a heat-bath with a much colder effective temperature determined by the electron's thermal spin polarization. By accurately determining the full spin Hamiltonian and performing realistic algorithmic simulations, we show that an experimental demonstration of heat-bath algorithmic cooling beyond the Shannon bound is feasible in both 3-qubit and 5-qubit variants of this spin system. Similar techniques could be useful for polarizing nuclei in molecular or crystalline systems that allow for non-equilibrium optical polarization of the electron spin. 

\end{abstract}
\maketitle
%
%
\section{Introduction}
Quantum Error Correction (QEC) is a critical tool for protecting quantum information against the imperfections of realistic devices and to scale quantum processors up to many qubits. Although a well developed theory exists \cite{knill1997theory,knill1998resilient,preskill1998reliable,knill2005quantum,aliferis2007accuracy,gottesman1997stabilizer} and experimental realizations at the several qubit level are recently emerging \cite{NVQEC,SCQEC}, there remain challenges for many potential implementations. Nuclear Magnetic Resonance (NMR) quantum information processing has demonstrated a high degree of quantum control \cite{knill2000algorithmic,negrevergne2006benchmarking,ryan2009randomized} and the ability to efficiently characterize the noise, both intrinsic and extrinsic, that affects the fidelity of a quantum process \cite{emerson2007symmetrized}. One piece has however been missing: the ability to efficiently polarize nuclear spin qubits on demand. The threshold theorem for quantum computation tells us that a quantum circuit can be simulated with arbitrarily high precision using a polynomial amount of resources as long as the error per gate $p$ is below a certain threshold value $p_{th}$ \cite{KLM}. The theorem relies on assuming that the ancilla qubits are in pure states at the beginning of each cycle of fault-tolerant QEC. For example, the first layer of concatenation of QEC typically reduces the effective error rate from $p$ to $cp^2$, where $c$ is a system-dependent constant. However, this theoretical gain is not generally achieved for the impure ancilla qubits characteristic of real implementations \cite{BenMixedQEC}. Thus, an efficient and experimentally feasible method for cooling qubits to high purity prior to each QEC cycle is desirable for all circuit implementations, including those based on nuclear spins.\\
\indent Heat bath algorithmic cooling (HBAC) \cite{AC_Boykin,AC_Fernandez,AC_Florian,AC_Kaye,PPA} is an efficient method for extracting entropy from a set of system qubits, allowing qubits to be cooled below the bath temperature (i.e. beyond the closed-system, or Shannon, bound). Solid state NMR experiments have demonstrated a sufficient level of coherent control to execute multiple rounds of algorithmic cooling, leading to spin polarizations exceeding the thermal polarization \cite{baugh2005experimental,AC_Ryan}. However, in a typical NMR setup, very low spin polarization at thermal equilibrium will require highly precise control over tens of nuclear spin qubits in order to polarize one ancilla qubit to order unity. Here, we explore the use of an electron spin to assist in the HBAC protocol. Due to its much larger gyromagnetic ratio compared to nuclei, the electronic thermal spin polarization is about $10^3$ times larger, and spin-lattice relaxation rates scale by a similar factor. Exploiting the electronic spin-lattice relaxation as a reset operation, the electron can connect a set of nuclear spins to heat bath with an effective temperature much lower than the equilibrium nuclear spin temperature. \\
\indent The stable malonyl radical ${\rm \dot{C}H(COOH)_2}$ has been extensively studied by electron spin resonance (ESR) studies of gamma-irradiated single crystals. The hyperfine tensors of the $\alpha$-proton and of the $^{13}$C-labeled methylene carbon were previously published \cite{cole1959electron,cole1961hyperfine,horsfield1961electron}. However, the hyperfine tensors of the $^{13}$C-labeled carboxyl carbons have not yet been reported. Including these two carboxyl carbons and assuming they are spectroscopically distinct, the molecule can in principle realize a 5-qubit ensemble quantum information processor, with 1 electron and 4 nuclear spins. In this paper we study the per-$^{13}$C-labeled radical and determine the hyperfine tensors of the carboxyl carbons via ESR and electron-nuclear double resonance (ENDOR) experiments on single crystal samples at room temperature. The carboxyl tensors are in fact different owing to different dihedral angles of COOH relative to the C-C-C plane to accommodate the hydrogen bonding network; the carboxyl group with a larger dihedral angle has a slightly weaker carbon hyperfine coupling. The previously published tensors describing the electronic $g$-factor and proton and methylene carbon hyperfine couplings are confirmed by our measurements. Given full knowledge of the spin Hamiltonian, we determine the optimal magnetic field orientations for carrying out quantum algorithms, focusing on HBAC as the example of interest. We consider two distinct methods for obtaining high fidelity coherent control, (i) use of anisotropic hyperfine interactions \cite{khaneja2007PRA,hodges2008universal,zhang2011coherent} facilitated by GRAPE \cite{khaneja2005optimal} pulses, and (ii) pulsed ENDOR control sequences. For brevity, we refer to approach (i) as Anisotropic Hyperfine Control (AHC). The two control schemes have very different criteria for choosing a suitable magnetic field orientation, given the desire to reduce the durations and optimize the fidelities of the quantum operations. Simulations of HBAC with realistic values of electronic spin-lattice relaxation time ($T_1$), intrinsic dephasing time ($T_2$) and ensemble dephasing time ($T^*_2$) are carried out for both the 3-qubit (methylene $^{13}$C) and 5-qubit (per-$^{13}$C labeled) versions. The results indicate that experimental cooling of nuclear spins beyond the thermal electron spin temperature, while challenging, is feasible in this quantum processor. These techniques can be applied to malonic acid or similar radical molecules at very low temperatures, or to optically polarizable systems such as NV centers in diamond \cite{loubser1978electron,doherty2011negatively,doherty2012theory} or molecules with photoexcited triplet states \cite{Akhtar2012PET} at room temperature, to reach the nuclear spin polarizations of order unity useful for QEC. \\
\indent The reminder of the paper is organized as follows. Section~\ref{Sample} provides relevant information about the properties and preparation of the single crystal samples. In section~\ref{Hamiltonian}, we review the extraction of $g$-factor and hyperfine tensors from orientation studies of ESR and ENDOR transitions, and report the tensors we determined including the $^{13}$C carboxyl tensors. Section~\ref{CrystalOrientation} presents  criteria for choosing optimal orientations of the magnetic field with respect to the crystalline axes, to facilitate control using either anisotropic hyperfine or pulsed ENDOR techniques. Realistic simulations of HBAC protocols for both 3 and 5 qubits are described in section~\ref{Simulation}. Conclusions are drawn in section~\ref{Conclusion}. 

\section{Sample preparation}
\label{Sample}
\indent Malonic acid, CH$_2$(COOH)$_2$, crystalizes with a triclinic unit cell and belongs to the ${\rm P}\bar1$ space group \cite{goedkoop1957crystal,jagannathan1994refinement} at temperatures above $47$ K. At lower temperatures, a structural phase transition occurs that has been discussed previously \cite{krzystek1995ednmr,fukai1991thermodynamic,mccalley1993endor}. Above $47$ K, there are two molecules per unit cell related by inversion symmetry, making them magnetically equivalent. We denote the methylene and carboxylic carbons as C$_m$ and C$_{1,2}$, respectively. A schematic of the radical, obtained by removing one of the methylene protons, is shown in Fig. 1. The unpaired electron is of $p$-orbital character \cite{cole1961hyperfine,mcconnell1960radiation,morton1964electron}. Malonic acid powder with all possible $^{13}$C isotopic labelling configurations was purchased either from Sigma Aldrich or Cambridge Isotopes. Single crystals were grown by slow evaporation from aqueous solutions at room temperature. To form radicals, crystals were irradiated to a dose of about $2$ kGy at room temperature with $\gamma$-rays from a cobalt-60 pencil. Annealing at $60\degree$ for 12-15 hours following the irradiation suppresses ESR signals from all other radical species except for the most stable radical, ${\rm \dot{C}H(COOH)_2}$ \cite{mcconnell1960radiation}. Depending on the $^{13}$C labelling configuration, 2, 3, 4 or 5 qubit samples are obtained as (e-H), (e-H-C$_m$), (e-H-C$_{1,2}$) and (e-H-C$_m$-C$_{1,2}$) respectively, where e denotes the electron spin. The $\beta$-protons of the carboxyl groups contribute to ESR line broadening and in some orientations can give rise to observable splittings \cite{sagstuen2000weakly}. However, due to weak hyperfine coupling ($<6$ MHz isotropic coupling \cite{sagstuen2000weakly}) they are not useful as qubits.

\begin{figure}[h]
\centering
\includegraphics[width=0.46\textwidth]{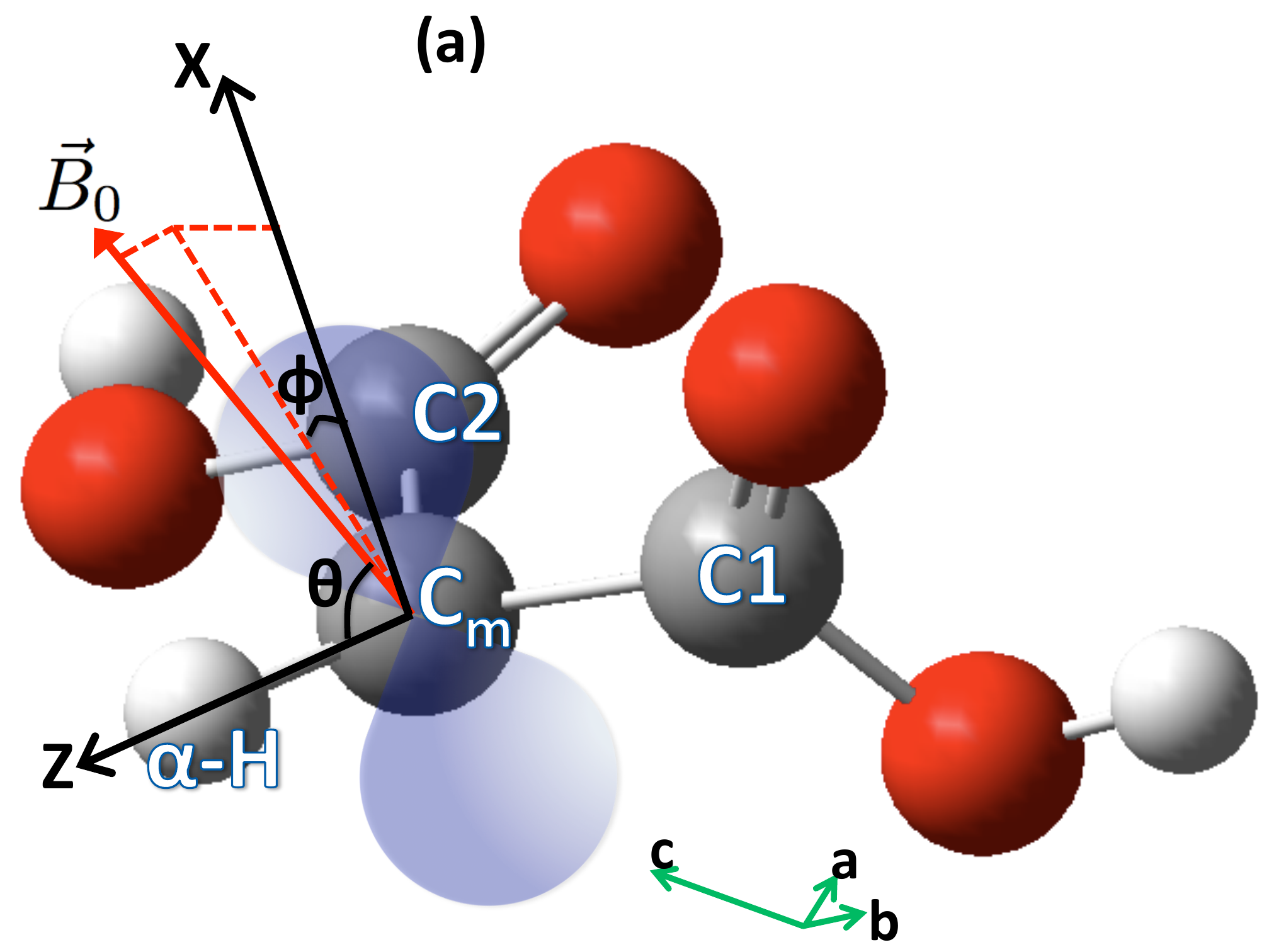}
\hspace{-1mm}\includegraphics[width=0.28\textwidth]{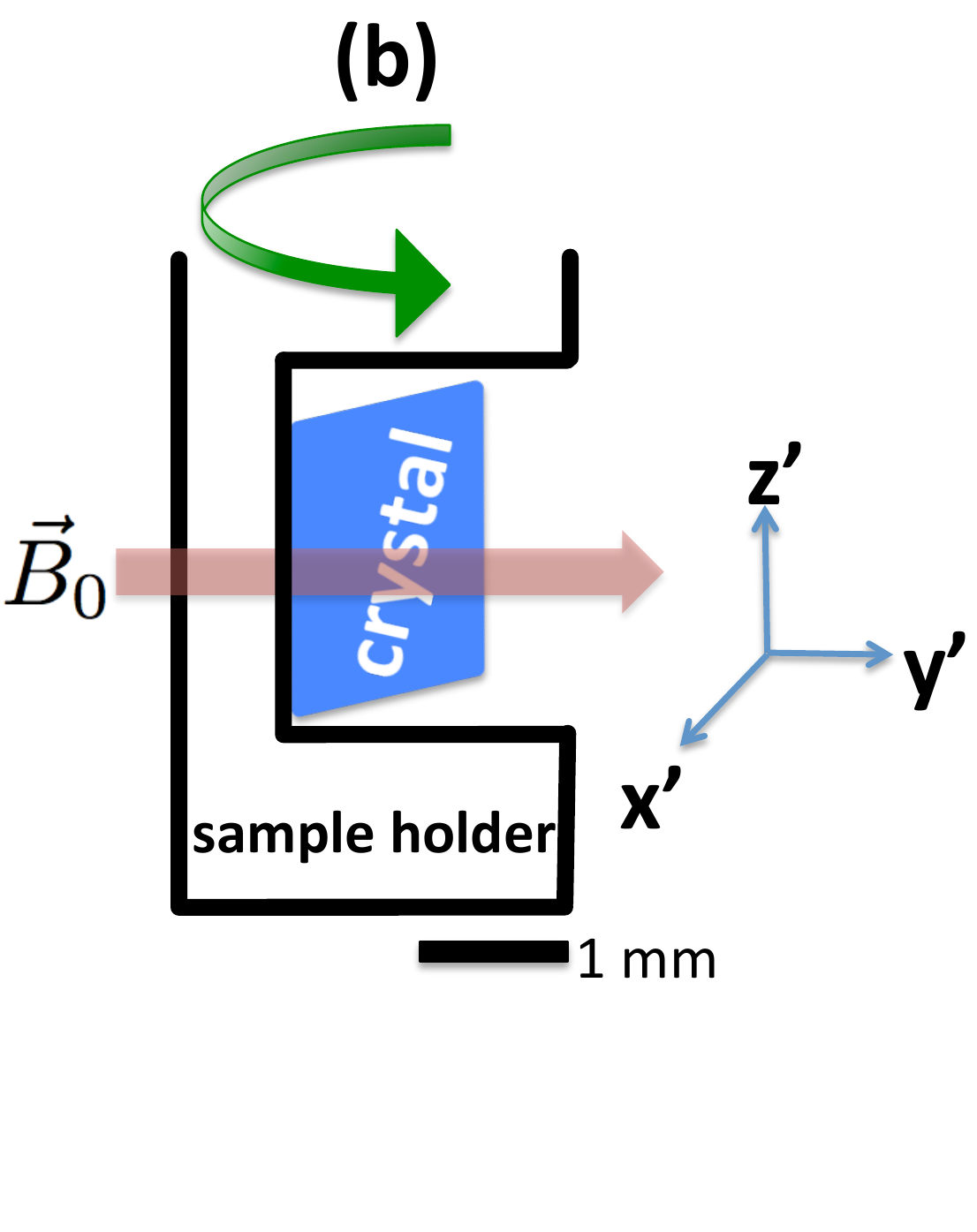}
\caption{\label{molecule}(a) Molecular structure of $\dot{\text{C}}$H(COOH)$_2$ with unpaired electron density distribution schematically represented by the blue shaded region. $x$ and $z$ are two principal axes of the $\alpha$-proton hyperfine tensor, with $z$ along the C$_m$--$\alpha$-proton interatomic vector and $x$ along the cylindrical symmetry axis of the electronic $p$-orbital. The $y$ axis, not shown, is nearly parallel to the C$_1$--C$_2$ interatomic vector. The direction of the static field $\vec{B}_0$ can be described using polar angle $\theta$ and azimuthal angle $\phi$ in the principal axis system of the $\alpha$-proton hyperfine tensor. The directions of the crystallographic axes $a$, $b$ and $c$ with respect to the molecular structure are illustrated by the green axes \cite{kang2003electronic,goedkoop1957crystal}. $c$ lies very close to the $y$ axis of the $\alpha$-proton principal axis system. The direction cosines of $a$, $b$ and $c$ in the $\alpha$-proton principal axis system are (0.1426, -0.6588, 0.7387), (-0.9729, -0.1888, 0.1335) and (0.0222, 0.9969, -0.0752), respectively. (b) Schematic of the crystal mounted on a sample holder for the orientation study (see section~\ref{CWESR}). Primed axes ($x'$, $y'$ and $z'$) indicate the lab frame in which measurements are taken. The curved green arrow indicates rotation of the crystal for orientation studies.}
\end{figure}

%
%
\section{Spin Hamiltonian}
\label{Hamiltonian}
The spin Hamiltonian of malonic acid with $K$ nuclei contains electron Zeeman, electron-nuclear hyperfine, and nuclear Zeeman terms as following:
\begin{equation}
\label{eq:HamModel1}
\mathcal{\hat{H}} = \mathcal{\hat{H}}_{e}+\mathcal{\hat{H}}_{hf}+\mathcal{\hat{H}}_{n}=\mu_B g_{\alpha\beta}B_{0\alpha}\hat{S}_\beta + \sum_{n=1}^K\left(2\pi A^n_{\alpha\beta}\hat{S}_\alpha\hat{I}^n_\beta -\gamma_{n} \hat{I}^n_\alpha B_{0\alpha}\right).
\end{equation}
Here $\alpha, \beta \in \lbrace x, y, z \rbrace$, and the repeated index implies summation over all the values of the index (similar to the Einstein summation convention, except all indices appear as lower indices). We set $\hbar = 1$ so that all Hamiltonians will appear in angular frequency units. $\hat{S}_{\alpha,\beta}$ and $\hat{I}_{\alpha,\beta}$ represent electron and nuclear spin operators, $\mu_B$ is Bohr magneton, $\vec{B}_0=(B_{0x},B_{0y},B_{0z})$ is external magnetic field and $\gamma_n$ is the gyromagnetic ratio for nuclear spin $n$. The second rank tensors $\tens{g}$ whose elements are $g_{\alpha\beta}$, and $\tens{A^n}$ whose elements are $A^n_{\alpha\beta}$, are the electron g-tensor and the hyperfine tensor describing coupling to nuclear spin $n$. In X-band ($\sim 10$ GHz) ESR, the electron Zeeman interaction is the dominant Hamiltonian term. The nuclear dipole-dipole interaction is neglected since it is typically at least two orders of magnitude smaller than the hyperfine interaction. When the nuclear Zeeman and hyperfine interaction energies are comparable and much smaller than the electron Zeeman energy, taking $\vec{B}_0=B_0\vec{z}$ the secular spin Hamiltonian can be written:
\begin{equation}
\label{HamApp}
\mathcal{\hat{H}}=\omega_{S}\hat{S}_z-\sum_{n=1}^K\omega_{I}^n\hat{I}_z^n
+2\pi \hat{S}_z\sum_{n=1}^K\left(A^n_{zz}\hat{I}_z^n+B^n\hat{I}_x^n\right).
\end{equation}
Here, $\omega_S = g_{zz}\mu_B B_0$ is the electron Larmor frequency, $\omega^n_I = \gamma_n B_0$ is the nuclear Larmor frequency, and $B^n = \sqrt{(A^n_{zx})^2+(A^n_{zy})^2}$. 

\subsection{Tensor extraction from ESR and ENDOR data}
\label{Method}
In this section, we briefly review the Hamiltonian determination method explained in \cite{atherton1993principles}.
First, we will consider the procedure to determine the $g$-tensor.
For a static magnetic field $\vec{B}_0=B_0(l_x,l_y,l_z)$ ($l_x$,$l_y$ and $l_z$ are direction cosines of the magnetic field in the axes of the crystal, and $l_x^2+l_y^2+l_z^2=1$), the electron Zeeman energy levels can be rewritten as
\begin{equation}
E_{\pm}=\pm\frac{1}{2}\mu_B B_0(l_\alpha\Gamma_{\alpha\beta}l_\beta)^{1/2},\label{EZ}
\end{equation}
where we introduce $\tens{\Gamma}=\tens{g}\cdot\tens{g}$.
In the absence of nuclear spins, the magnetic resonance condition $\Delta E = h\nu=\mu_{B}gB_0$ leads to
\begin{equation}
g=\frac{h\nu}{\mu_{B}B_0}=\frac{\Delta E}{\mu_{B}B_0}=\frac{E_{+}-E_{-}}{\mu_{B}B_0}=(l_\alpha\Gamma_{\alpha\beta}l_\beta)^{1/2},\label{resonancce}
\end{equation}
where $g$ is the experimentally observed value, $\Delta E$ is the difference between $E_{+}$ and $E_{-}$,  and $\nu$ is the frequency of the microwave (mw) field. For $\vec{B}_0=B_0(\cos{\theta},\sin{\theta},0)$, using the symmetry $\Gamma_{\alpha\beta}=\Gamma_{\beta\alpha}$, the dependence of $g$ on $\theta$ can be written as
\begin{equation}
(g(\theta))^{2}=l_\alpha\Gamma_{\alpha\beta}l_\beta=\Gamma_{xx}\cos^{2}\theta+\Gamma_{yy}\sin^{2}\theta+2\Gamma_{xy}\cos{\theta}\sin{\theta}.\label{depend}
\end{equation}

By measuring $g(\theta)$ at various $\theta$ and solving Eq. (\ref{depend}), $\Gamma_{xx}$, $\Gamma_{yy}$ and $\Gamma_{xy}$ can be determined. Similarly, by orienting $\vec{B}_0$ within the $x$-$z$ and $y$-$z$ planes and measuring $g(\theta)$ for various $\theta$ (where $\theta$ is the angle between $\vec{B}$ and $z$, or between $\vec{B}$ and $y$), $\Gamma_{zz}$, $\Gamma_{xz}$ and $\Gamma_{yz}$ can be determined. Therefore full knowledge of $\tens{\Gamma}$ can be obtained.

Now we consider the electron-nuclear coupled spin Hamiltonian which contains $\mathcal{\hat{H}}_{e}$, $\mathcal{\hat{H}}_{hf}$ and $\mathcal{\hat{H}}_{n}$. In this case, the 1st order approximation of the energy levels of the system are
\begin{align}
 E_{M_SM_I}=&g\mu_BB_0M_S+\sum_{n=1}^K\{\frac{4\pi^2}{g^2}(l_\alpha\Gamma_{\alpha\beta}^{A^n}l_\beta)+\frac{\gamma_n^2B_0^2}{M_S^2}\nonumber\\&-\frac{4\pi}{gM_S}\gamma_n B_0(l_\alpha(\tens{g}\cdot\tens{A^n})_{\alpha\beta}l_\beta)\}^{1/2}M_{I}^nM_S,\label{enerylevel}
\end{align}
where $\tens{\Gamma^A}=\tens{g}\cdot \tens{A}\cdot \tens{A}\cdot \tens{g}$, and  $M_S$ and $M_I^n$ are quantum numbers for the electron spin and the $n^{\text{th}}$ nuclear spin, respectively. 
All ESR transitions can be classified into two types: allowed and forbidden transitions. Allowed transitions correspond to electron-only spin flips, while forbidden transitions correspond to the flips of electron spin along with one or more nuclear spins. In the absence of anisotropic hyperfine coupling (the terms with $B^n$ coefficient in Eq. (\ref{HamApp})), forbidden transitions are completely suppressed and produce no observable ESR signal. When $B^n \neq 0$, these transitions can lead to signals comparable to, but typically smaller than, the allowed transitions. For the purpose of hyperfine tensor determination, it suffices to track only allowed ESR transitions. The nuclear Zeeman energy is generally much smaller than hyperfine coupling strength and is usually omitted in the traditional process of determining hyperfine tensors using ESR spectra. However when we determined hyperfine tensors in this work, after the traditional process we implemented an optimization process which takes the nuclear Zeeman energy into consideration. We will discuss this further in Section~\ref{CWESR}. In the present section we will stick to the traditional process. Therefore, after omitting the nuclear Zeeman energy, the ESR allowed transition energy gap in the 1 electron-1 nucleus system is 
\begin{align}
\Delta E=g\mu_BB_0+\frac{2\pi}{g}(l_\alpha\Gamma_{\alpha\beta}^{A}l_\beta)^{1/2}M_{I},
\end{align}
where the nuclear spin quantum numbers $M_{I}=\pm 1/2$ represent spin up and spin down states, respectively. The ESR spectrum then splits into two distinct peaks separated by a frequency $A=g^{-1}(l_\alpha\Gamma_{\alpha\beta}^{A}l_\beta)^{1/2}$. Similar to Eq. (\ref{depend}), when the static magnetic field is $\vec{B}_0=B_0(\cos{\theta},\sin{\theta},0)$ we have
\begin{align}
(g(\theta)A(\theta))^2=\Gamma^A_{xx}\cos^{2}\theta+\Gamma^A_{yy}\sin^{2}\theta+2\Gamma^A_{xy}\cos{\theta}\sin{\theta},\label{dependA}
\end{align}
and, as before, expressions for $\vec{B}_0$ in the $x$-$z$ and $y$-$z$ planes relate the other tensor components of $\tens{\Gamma^A}$ to $(gA)^2$. 

Another method to determine hyperfine tensors is to make use of ENDOR experiments. For ENDOR transitions, which correspond to $\Delta M_S=0$ and $\Delta M_I=\pm 1$, the energy gap in the 1 electron-1 nucleus system can also be readily obtained from Eq. (\ref{enerylevel}) and its square is  given by
\begin{equation}
\Delta E^2=\gamma^2B_0^2+\frac{4\pi^2M_S^2}{g^2}(l_\alpha\Gamma_{\alpha\beta}^{A}l_\beta)-\frac{4\pi M_S\gamma B_0}{g}(l_\alpha(\tens{g}\cdot\tens{A})_{\alpha\beta}l_\beta).\label{ENDOR1}
\end{equation}
Therefore, the ENDOR spectral peaks also split into two for $M_S=\pm 1/2$. Supposing that the observed ENDOR frequencies are $\nu_{\pm}$ for $M_S=\pm 1/2$, Eq. (\ref{ENDOR1}) can be rewritten as
\begin{equation}
\nu_{\pm}^2=\nu_I^2+\frac{1}{4g^2}(l_\alpha\Gamma_{\alpha\beta}^{A}l_\beta)\mp\frac{\nu_I}{g}(l_\alpha(\tens{g}\cdot\tens{A})_{\alpha\beta}l_\beta),\label{ENDOR}
\end{equation}
where $\nu_I=\gamma B_0/2\pi$. An equation similar to Eqs.(\ref{depend}) and (\ref{dependA}) can be obtained when the static magnetic field is $\vec{B}_0=B_0(\cos{\theta},\sin{\theta},0)$,
\begin{equation}
\frac{g}{2\nu_I}(\nu_-^2 -\nu_+^2)=(\tens{g}\cdot \tens{A})_{xx}\cos^{2}\theta+(\tens{g}\cdot \tens{A})_{yy}\sin^{2}\theta+2(\tens{g}\cdot \tens{A})_{xy}\cos{\theta}\sin{\theta}.\label{dependE}
\end{equation}
Therefore, the magnitude of $\tens{A}$ can be extracted by measuring the orientation dependence of $g(\nu_-^2 -\nu_+^2)/2\nu_I$ in three distinct rotation planes.

\subsection{Continuous-wave ESR results}
\label{CWESR}
The tensor extraction model relies on two assumptions: the three planes of measurement are mutually non-parallel and the axis of the rotation to go from one to another belongs to both planes. The orientation experiments were designed to ensure that both assumptions are satisfied. In each of the three non-parallel planes, we took 24, 9 and 46 measurements with 8$^{\circ}$, 20$^{\circ}$ and 4$^{\circ}$ angle steps for the methylene-labeled, carboxyl-labeled and fully-labeled MA samples, respectively. The methylene-labeled data was used to extract the $\alpha$-$^{1}$H and $^{13}$C$_m$ tensors; the carboxyl-labeled data was used to extract the $^{13}$C$_{1,2}$ (average) tensor, and the fully-labeled data was used to confirm all the tensors by fitting with simulated spectra. As discussed previously in Section~ \ref{Method}, we need only to detect the spectral positions of the allowed transitions, which can be distinguished from forbidden transition peaks because of their larger intensities. Each measured spectrum is fit to a set of allowed transition peaks defined by their amplitude, frequency and a common line width. \\
\indent The peak positions for all measurements in one plane give a set of trajectories for that plane, from which we obtained the dependence of $g$ and $A^n$ on $\theta$. Doing this in all three planes allows extraction of the $g$-factor tensor $\tens{g}$ and hyperfine tensors $\tens{A^n}$ by solving Eqs. (\ref{depend}) and (\ref{dependA}). After this, we used the method of least square fitting to optimize $\tens{g}$ and $\tens{A^n}$ by minimizing the difference between the experimental peak trajectories and the simulated trajectories. It should be noted that in X-band ESR the nuclear Zeeman energy for $^1$H is about $15$ MHz, which is comparable to the hyperfine coupling of $\alpha$-$^1$H in some orientations. Therefore, in order to obtain more accurate $\tens{A^n}$ tensors,  we took into account the nuclear Zeeman energy when generating the simulated peak trajectories in the optimization procedure. The only noticeable difference from including this energy was for the $\alpha$-$^1$H, as expected. \\
\indent The $\tens{g}$, $\tens{A^H}$, $\tens{A^{Cm}}$ and $\tens{A^{C_{1,2}}}$ tensors were determined from continuous (CW) ESR spectra as described above, and are listed in Table~\ref{ESRresult}. The hyperfine tensors of $\rm C_1$ and $\rm C_2$ are similar and require ENDOR measurements to be distinguished (see next section). From the ESR data we obtained an average hyperfine tensor describing $\rm C_1$ and $\rm C_2$, denoted as $\tens{A^{C_{1,2}}}$. Negative signs of the principal values for $\tens{A^{C_{1,2}}}$ were determined by a quantum chemical calculation using the Gaussian program~\cite{g09}. Similar to the case of $\alpha$-$^1$H \cite{cole1961hyperfine,mcconnell1960radiation,morton1964electron}, the negative principal values are due to a negative spin density on $\text{C}_1$ and $\text{C}_2$.

\begin{table}[h]\footnotesize
\renewcommand{\arraystretch}{1.5}
\centering
\begin{tabular}{cc||C{1.5cm}|C{1.5cm}|C{1.5cm}|}
\cline{3-5}
&\multicolumn{1}{c|}{} &\multicolumn{3}{c|}{Direction cosines to $\tens{A^H}$ principal axis} \\ \cline{3-5}
& \multicolumn{1}{c|}{}& X & Y & Z \\ \cline{1-5}
\multicolumn{1}{|c|}{g$_{\text{xx}}$} & 2.00250 $\pm$ 0.00038 & -0.1657 & 0.9779 & 0.1272 \\
\multicolumn{1}{|c|}{g$_{\text{yy}}$} & 2.00373 $\pm$ 0.00037 & -0.9811 & -0.1766 & 0.0797\\
\multicolumn{1}{|c|}{g$_{\text{zz}}$} & 2.00417 $\pm$ 0.00036 & 0.1004 & -0.1115 & 0.9887 \\ \cline{1-5}
\multicolumn{1}{|c|}{A$^{\text{H}}_{\text{xx}}$} & -26.6 $\pm$ 2.8 & 1 & 0 & 0 \\
\multicolumn{1}{|c|}{A$^{\text{H}}_{\text{yy}}$} & -56.0 $\pm$ 0.7 & 0 & 1 & 0 \\
\multicolumn{1}{|c|}{A$^{\text{H}}_{\text{zz}}$} & -91.5 $\pm$ 0.6 & 0 & 0 & 1 \\ \cline{1-5}
\multicolumn{1}{|c|}{A$^{\text{Cm}}_{\text{xx}}$} & 24.5 $\pm$ 1.0 & 0.0696 & -0.0019 & 0.9976 \\
\multicolumn{1}{|c|}{A$^{\text{Cm}}_{\text{yy}}$} & 43.0 $\pm$ 1.3 & 0.9962 & 0.0530 & -0.0694 \\
\multicolumn{1}{|c|}{A$^{\text{Cm}}_{\text{zz}}$} & 212.3 $\pm$ 0.6 & -0.0528 & 0.9986 & 0.0056 \\ \cline{1-5}
\multicolumn{1}{|c|}{A$^{\text{C}_{1,2}}_{\text{xx}}$} & -36.1 $\pm$ 0.3 & 0.0627 & 0.0082 & 0.9980 \\
\multicolumn{1}{|c|}{A$^{\text{C}_{1,2}}_{\text{yy}}$} & -39.3 $\pm$ 0.3 & 0.9805 & -0.1870 & -0.0601 \\
\multicolumn{1}{|c|}{A$^{\text{C}_{1,2}}_{\text{zz}}$} & -40.6 $\pm$ 0.3 & 0.1862 & 0.9823 & -0.0198 \\ \cline{1-5}
\end{tabular}
\renewcommand{\arraystretch}{1}
\caption{\label{ESRresult} Electronic g-factor and hyperfine coupling tensors determined from CW ESR measurements (principal hyperfine values are given in MHz). Principal values are given in the left column, while direction cosines relative to the principle axis system of the $\alpha$-proton are given in the three right columns. The uncertainties reflect a $90\%$ confidence interval. The tensor $\tens{A^{C_{1,2}}}$ gives an estimate of the average of $\tens{A^{C_1}}$ and $\tens{A^{C_2}}$.}
\end{table}

\begin{figure}[h]
	\centering	
	\includegraphics[width=0.6\textwidth]{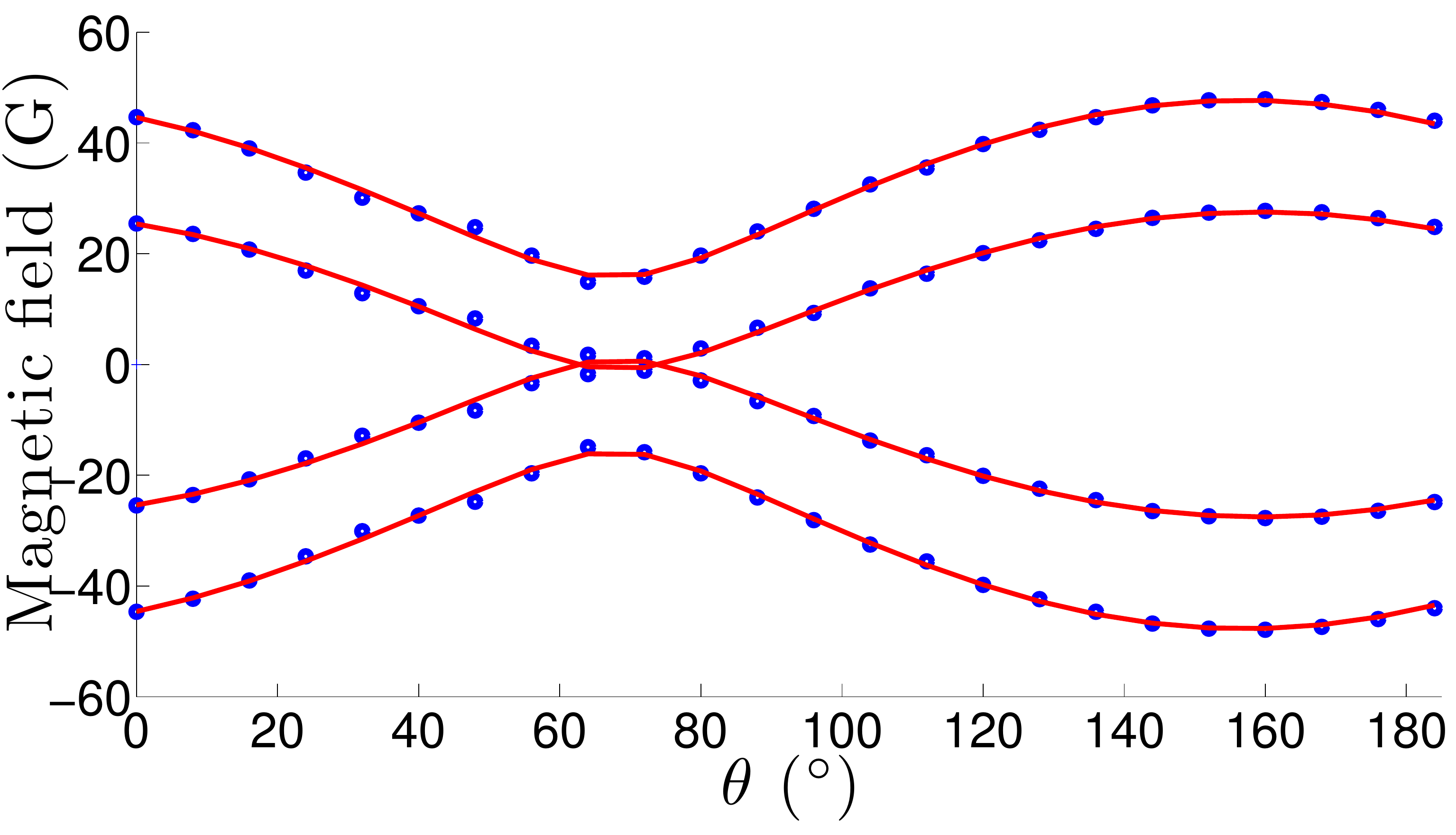}
	\caption{\label{central1} Comparison between the simulated ESR peak trajectories (red lines) and the experimental data (blue dots) of the methylene $^{13}$C-labeled sample in one of the planes measured. The direction cosines of the normal of this plane are (-0.3376   0.9405    0.0391) in the principal axis system of $\tens{A^H}$.}
\end{figure}
\begin{figure}[h]
	\centering
	\includegraphics[width=0.6\textwidth]{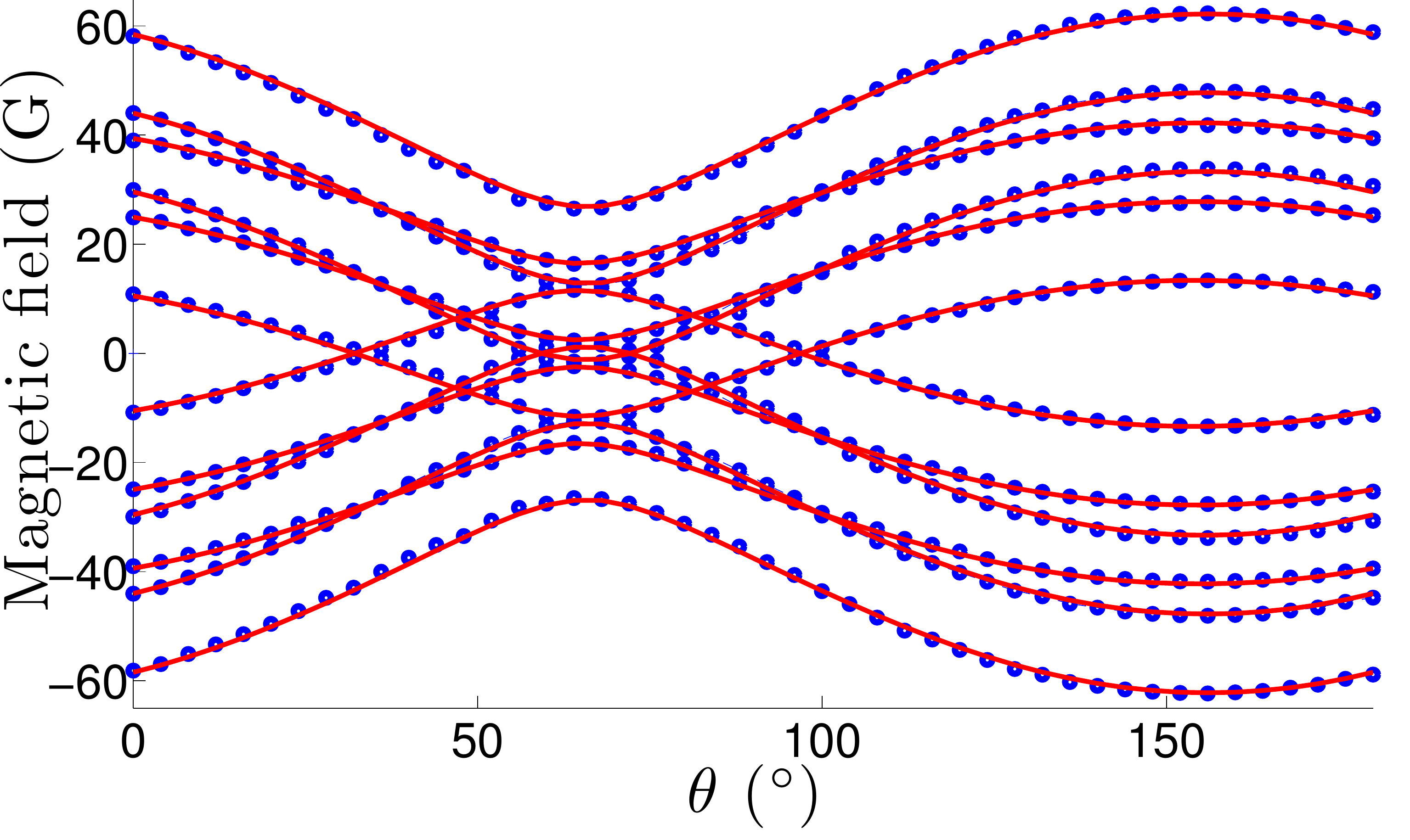}
	\caption{\label{fully1} Comparison between the simulated ESR peak trajectories (red lines) and the experimental data (blue dots) of the per-$^{13}$C-labeled sample in one of the planes measured. The direction cosines of the normal of this plane are (-0.3678   0.9287   -0.0483) in the principal axis system of $\tens{A^H}$.}
\end{figure}

The results for $\tens{g}$, $\tens{A^H}$ and $\tens{A^{Cm}}$ are consistent with the published results in \cite{cole1959electron,cole1961hyperfine,horsfield1961electron,mccalley1993endor,mcconnell1960radiation,kang2003electronic}. The simulated ESR peak trajectories generated from the tensors in Table \ref{ESRresult} are in excellent agreement with the experimental data, as the examples in Figs. \ref{central1} and \ref{fully1} demonstrate. 

\subsection{Continuous-wave ENDOR results} 
\begin{figure}[h]
\centering
\includegraphics[width=0.6\textwidth]{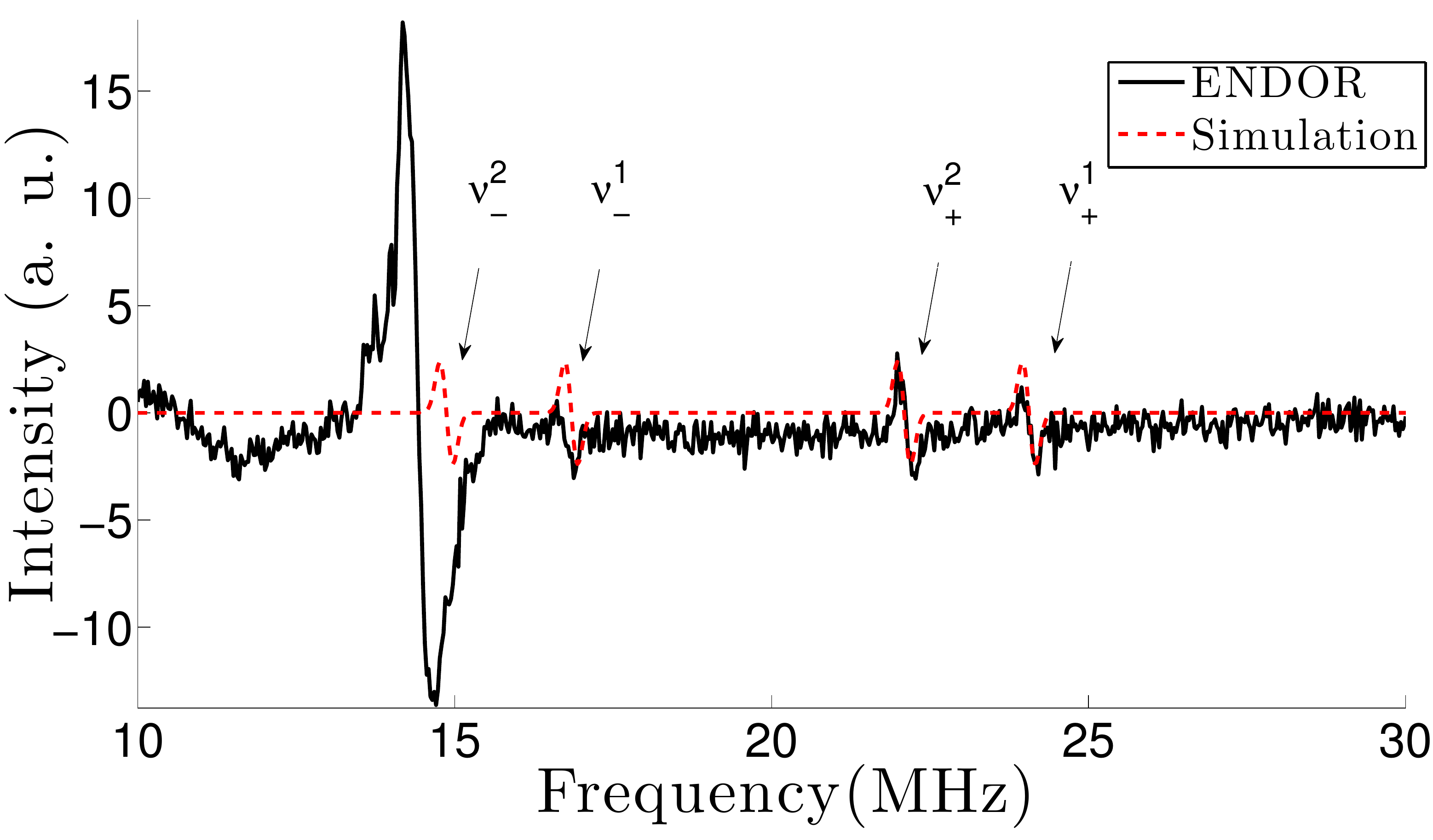}
	\caption{\label{Endorspec} Comparison between the simulated ENDOR spectrum (red line)  and the experimental spectrum (black line) of the carboxyl $^{13}$C-labeled sample. The orientation of magnetic field is (0.6156, -0.7179, -0.3249) in the principal axis system of $\tens{A^H}$. In the area around $14.5$ MHz, there are large peaks which come from the distant $^1\text{H}$ spins that have small hyperfine couplings with the electron radical. The ENDOR peak for $\text{C}_2$ with the frequency $\nu_-^2=14.9$ MHz is obscured by these $^1\text{H}$ peaks. The other three ENDOR peaks for $\text{C}_1$ and $\text{C}_2$ with frequencies  $\nu_+^1=24.1$ MHz, $\nu_+^2=22.1$ MHz and $\nu_-^1=16.8$ MHz are clear.}
\end{figure}

In this section we describe the use of ENDOR, which has higher spectral resolution than ESR, to extract the distinct tensors describing $\tens{A^{C_1}}$ and $\tens{A^{C_2}}$. From Eq. (\ref{ENDOR}) we see that for each nucleus there are two resonant frequencies $\nu_-$ and $\nu_+$. However, in experiments we found that one or sometimes two of the four peaks of $\text{C}_1$ and $\text{C}_2$ were obscured by the large peaks around $14.5$ MHz that come from $^1\text{H}$ spins which have small hyperfine couplings with the electron radical (Fig. \ref{Endorspec}). The two higher frequency peaks (in the range of $20-25$ MHz) were clearly resolved in all measured orientations. Judging from the sign of $\tens{A^{C_{1,2}}}$, we infer that these two peaks correspond to the $\nu_+^{\text{C}_1}$ and $\nu_+^{\text{C}_2}$ frequencies, with
\begin{equation}
\label{ENDORsingle}
(\nu_{+}^n)^2=(\nu_I^n)^2+\frac{1}{4g^2}(l_\alpha\Gamma_{\alpha\beta}^{A^n}l_\beta)-\frac{\nu_I^n}{g}(l_\alpha(\tens{g}\cdot\tens{A^n})_{\alpha\beta}l_\beta)=(\nu_I^n)^2+\frac{1}{g^2}(l_\alpha K_{\alpha\beta}^{A^n}l_\beta),
\end{equation}
where $n=1,2$ for $\text{C}_1$ and $\text{C}_2$ and $\tens{K^{A^n}}=\tens{\Gamma^{A^n}}/4-\nu_I^ng\tens{g}\cdot\tens{A^n}$. Using only the $\nu_+$ frequencies requires a different approach than directly using Eq. (\ref{dependE}). Instead of measuring the orientation dependence of $g((\nu_-^n)^2 -(\nu_+^n)^2)/2\nu_I^n$, we measured that of $g^2((\nu_{+}^n)^2-(\nu_I^n)^2)$ in three non-parallel planes to extract $\tens{K^{A^n}}$ using Eq. (\ref{ENDORsingle}), and then used $\tens{K^{A^n}}$ to calculate $\tens{A^{C_1}}$ and $\tens{A^{C_2}}$. In each plane, we took 9 measurements with 20$^{\circ}$ angle steps for the carboxyl-labeled MA sample. It should be mentioned that by using Eq. (\ref{ENDORsingle}) to obtain $\tens{A^{C_1}}$ and $\tens{A^{C_2}}$, we neglected the anisotropy of $\tens{g}$. For the hyperfine tensors, this approximation causes a difference in the principal values less than the experimental uncertainty of $\sim 0.1$ MHz. The experimentally determined $\tens{A^{C_1}}$ and $\tens{A^{C_2}}$ tensors are listed in Table \ref{Endorresult}.

\begin{table}[h]\footnotesize
\renewcommand{\arraystretch}{1.5}
\centering
\begin{tabular}{cc||C{1.5cm}|C{1.5cm}|C{1.5cm}|}
\cline{3-5}
& \multicolumn{1}{c|}{}&\multicolumn{3}{c|}{Direction cosines to $\tens{A^H}$ principal axis} \\ \cline{3-5}
& \multicolumn{1}{c|}{}& X & Y & Z \\ \cline{1-5}
\multicolumn{1}{|c|}{A$^{\text{C}_{1}}_{\text{xx}}$} & -37.0 $\pm$ 0.1 & -0.5310 & -0.0010 & 0.8474 \\
\multicolumn{1}{|c|}{A$^{\text{C}_{1}}_{\text{yy}}$} & -40.5 $\pm$ 0.1 & 0.8452 & -0.0724 & 0.5295 \\
\multicolumn{1}{|c|}{A$^{\text{C}_{1}}_{\text{zz}}$} & -43.6 $\pm$ 0.1 & 0.0608 & 0.9974 & 0.0392 \\ \cline{1-5}
\multicolumn{1}{|c|}{A$^{\text{C}_{2}}_{\text{xx}}$} & -34.0 $\pm$ 0.1 & -0.3477 & -0.0449 & -0.9365 \\
\multicolumn{1}{|c|}{A$^{\text{C}_{2}}_{\text{yy}}$} & -37.4 $\pm$ 0.1 & -0.2478 & 0.9677 & 0.0456 \\
\multicolumn{1}{|c|}{A$^{\text{C}_{2}}_{\text{zz}}$} & -39.6 $\pm$ 0.1 & -0.9043 & -0.2479 & 0.3476 \\ \cline{1-5}
\end{tabular}
\renewcommand{\arraystretch}{1}
\caption{\label{Endorresult} Hyperfine coupling tensors for $\tens{A^{C_1}}$ and $\tens{A^{C_2}}$ determined from CW ENDOR measurements (principal hyperfine values are given in MHz). Principal values are given in the left column, while direction cosines relative to the principle axis system of the alpha-proton are given in the three right columns. The uncertainties reflect a $90\%$ confidence interval.}
\end{table}

\begin{figure}[h]
\centering
\includegraphics[width=0.6\textwidth]{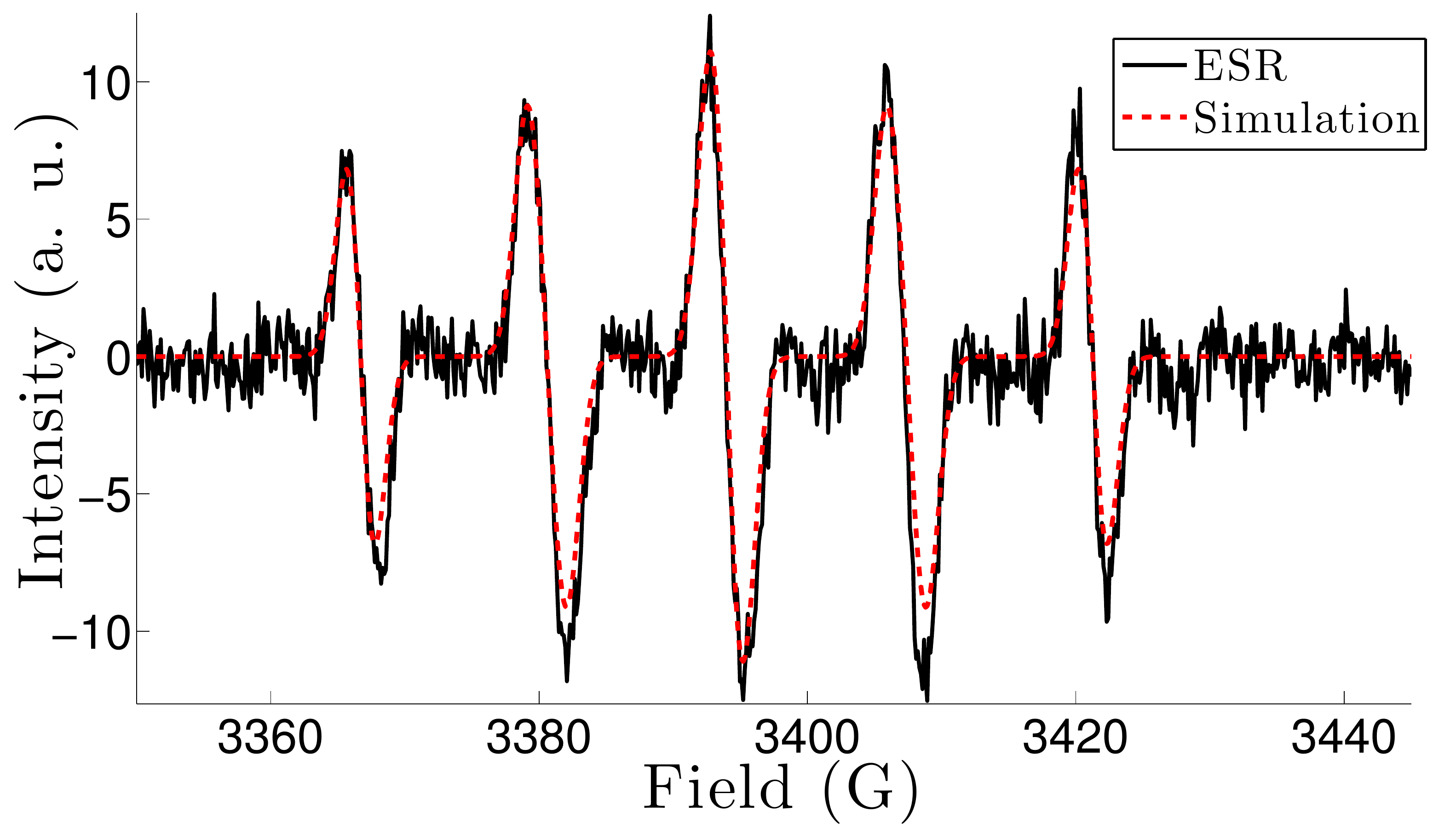}
	\caption{\label{Esrspec}Comparison between the simulated ESR spectrum (red line)  and the experimental spectrum (black line) of the carboxyl $^{13}$C-labeled sample. The orientation of magnetic field is (0.6156, -0.7179, -0.3249) in the principal axis system of $\tens{A^H}$. Using the tensors given in Table~\ref{Endorresult}, the ESR peak positions are reproduced well and the line broadening due to the hyperfine couplings of the two carboxyl carbons fits well in all measured orientations.}
\end{figure}

Figs. \ref{Endorspec} and \ref{Esrspec} show the experimental ENDOR and ESR spectra, together with the simulated spectra generated using the tensors given in Table \ref{Endorresult}, for the carboxyl-labeled sample when the orientation of the magnetic field is (0.6156    -0.7179   -0.3249) in the principal axis system of $\tens{A^H}$. In Fig. \ref{Endorspec}, we can see that in this orientation there are three observable peaks of $\text{C}_1$ and $\text{C}_2$, while the peak with the lowest frequency is obscured by the $^1\text{H}$ peaks around $14.5$ MHz. Excellent agreement between experiment and simulation can be seen in these examples, and a similar level of agreement was found for spectra in all measured orientations, indicating that the tensors in Table  \ref{Endorresult} are accurate.

%
%
\section{Optimizing magnetic field orientation for quantum control}
\label{CrystalOrientation}
\subsection{Orientation criteria for AHC}
\label{COAHC}
In the presence of $\vec{B}_0=B_0\vec{z}$, a hyperfine-coupled nuclear spin is quantized along the direction of an effective field
\begin{equation}
\vec{B}_{\text{eff}}=\left(B_0\pm \pi A_{zz}/\gamma\right)\vec{z}\pm \left(\pi B/\gamma\right)\vec{x},
\end{equation}
where $B=\sqrt{(A_{zx})^2+(A_{zy})^2}$ and the $\pm$ sign depends on whether the electron spin is parallel ($\kket{\uparrow}$) or anti-parallel ($\kket{\downarrow}$) to $\vec{B}_0$. The angle of $n^{\text{th}}$ nuclear spin quantization axis from $z$ is denoted as $\eta_{\uparrow}^n=\arctan\left(-B^n/(A^n_{zz}+\omega_I^n/\pi)\right)$ for electron spin $\kket{\uparrow}$ and $\eta_{\downarrow}^n=\arctan\left(-B^n/(A^n_{zz}-\omega_I^n/\pi)\right)$ for electron spin $\kket{\downarrow}$. \\
\indent A requirement for achieving high control fidelity is to implement gates on a time scale fast compared to the electronic dephasing time $T_2^{e}$. The electron spin control Hamiltonian in the rotating frame is $\mathcal{H}_c(t)=\omega_1(t)(\hat{S}_x \cos{(\phi(t))} + \hat{S}_y \sin{(\phi(t))}$, where $\omega_1$ and $\phi$ represent the microwave amplitude and phase, respectively. Consider $\omega_1 = $ constant, and $\phi=0$. Nuclear spin flips can occur via electron-nuclear flip-flip forbidden transitions driven at a rate $\omega_1\sin(\eta^n)/2$ where $\eta^n=(\eta^n_{\uparrow}-\eta^n_{\downarrow})/2$. The on-resonance allowed ESR transition is driven at $\omega_1\cos(\eta^n)/2$. Depending on the nuclear isotope and the dc field orientation, $\sin(\eta^n)/2$ can be very small, preventing efficient control of the nuclear spins via microwave excitation. Thus, we aim to find an orientation that yields the set $\{ A^n_{zz},B^n\}$ that maximizes the ratio of forbidden to allowed transition rates, $\tan(\eta^n)$. Secondly, all allowed and forbidden transition frequencies should be separated from each other by at least the ESR line width ($\sim 12$ MHz in the malonyl radical) in order to achieve universal control. 

For the optimal orientation search, $B_0$ is fixed at $2\pi\cdot 10$ GHz$/\gamma_e\approx 3568$ G (note $\gamma_e=g\mu_B$) and $\theta$ and $\phi$ (see Fig.~\ref{molecule}(a)) are varied at $1\degree$ angle steps. The maximum values of $\tan(\eta^n)$ are 1.3, 0.14, 0.016, and 0.017 for H, C$_m$, C$_1$, and C$_2$, respectively. The duration of a GRAPE optimal control \cite{khaneja2005optimal} pulse to accomplish an arbitrary unitary operation will scale as $1 / \tan(\eta^n)$. We have measured a nearly temperature-independent electron spin dephasing time $T^e_2 \sim 5$ $\mu$s in this system, due to hyperfine coupling with distant protons, and their mutual dipolar interaction induced dynamics. Given that the shortest GRAPE pulses for arbitrary gates involving H require $\sim 500$ ns, it is clear that high fidelity control of C$_1$, and C$_2$ via AHC is simply not possible in this system. Moreover, the 5-qubit system contains 80 ESR resonances over a maximum spectral range of $250$ MHz, consisting of 16 allowed and 64 forbidden transitions. Consequently, we find $\max_{\theta,\phi}\{\min(\vec{\delta})\}=1.4$ MHz, where the elements of $\vec{\delta}$ are the distances between any two ESR transitions and $\max_{\theta,\phi}\{\}$ indicates the maximization over all sets of $\{\theta,\phi\}$ with $1\degree$ angle step. In practice, it only makes sense therefore to focus on the 3-qubit sample (e-H-C$_m$) for optimizing AHC control. Nonetheless, for completeness we show in Table~\ref{AHCorientation} the best possible orientation for the 5-qubit system given these criteria. \\
\indent For the 3-qubit sample, there are 4 allowed transitions and 8 forbidden transitions. It is still not possible to find orientations where $\min(\vec{\delta})\ge 12$ MHz, rather we find that $\max_{\theta,\phi}\{\min(\vec{\delta})\}=7.6$ MHz. We focus on orientations that satisfy $\min(\vec{\delta})\ge 6$ MHz, and then choose an orientation in which $\tan(\eta^{H})$ and $\tan(\eta^{C_{m}})$ are both as large as possible. The forbidden transition rates at this orientation are 0.1418 and 0.1203 for H and C$_m$, respectively, and $\min(\vec{\delta})=6.1$ MHz. These results are summarized in the table below:
\begin{table}[h]
\renewcommand{\arraystretch}{1.2}
\centering
\begin{tabular}{c c|C{1cm}|C{1cm}|C{1cm}|C{1cm}| c}
\cline{3-6}
& &\multicolumn{4}{c|}{$A^n_{zz}$(MHz), $B^n$(MHz), $\tan(\eta^n)$}&  \\ \cline{2-7}
\multicolumn{1}{C{1cm}|}{}& $\theta,\phi$ & H & C$_m$ & C$_{1}$&C$_{2}$& \multicolumn{1}{C{1cm}|}{min$(\vec{\delta})$ (MHz)}\\ \cline{1-7}
\multicolumn{1}{|C{1cm}|}{5 qubit} & $54\degree,123\degree$ & -62.56, 23.67, 0.1948 & 121.23, 90.54, 0.0303 & -40.03, 3.27, 0.0161& -37.29, 1.63, 0.0093&\multicolumn{1}{C{1cm}|}{1.1}\\ \hhline{|=|=|=|=|=|=|=|}
\multicolumn{1}{|C{1cm}|}{3 qubit} & $29\degree$, $13\degree$& -76.60, 27.07, 0.1418 & 29.21, 17.78, 0.1203 & & &\multicolumn{1}{C{1cm}|}{6.1}\\ \cline{1-7}
\end{tabular}
\renewcommand{\arraystretch}{1}
\caption{\label{AHCorientation}Optimal orientations for AHC experiment with the fully-labeled (5 qubit) and C$_m$-labeled (3 qubit) malonyl radical, and corresponding values of hyperfine coupling constants $A^n_{zz}$ and $B^n$, and the quantity $\tan(\eta^n)$ characterizing the controllability of nuclear spin $n$. min$(\vec{\delta})$ represents the minimum separation between ESR transitions.}
\end{table}

\subsection{Orientation criteria for ENDOR}
\label{OCENDOR}
In the ENDOR control scheme, nuclear spin rotations are implemented directly by applying radio-frequency (RF) pulses on resonance with the nuclear transitions. Hence, $\tan(\eta)$ does not need to be large. In fact, it is desirable to minimize $\tan(\eta)$, because the electron spin-lattice relaxation process induces nuclear spin relaxation through the two-spin flip forbidden transitions. As before, the allowed ESR transitions must be separated by at least the ESR line width, and the NMR transitions must be separated by at least the NMR (ENDOR) linewidth. The latter is determined by the nuclear dephasing time $T^n_2 >> T^e_2$, and is therefore much narrower than the ESR line width. The disadvantage of ENDOR control is that the RF pulses are typically of order $\gamma_e / \gamma_n \sim 10^3$ times slower than the microwave control of electron spin. This severely limits the ability to perform arbitrary quantum algorithms within the electron $T^e_2$, however, in algorithmic cooling the electron spin is always in an eigenstate during the nuclear rotations so that electronic dephasing is not an issue. Since the hyperfine tensors of C$_1$ and C$_2$ are very similar, they cannot be separately addressed by microwave pulses, but only with RF pulses. At first glance, this seems to forbid selective swap gates between the electron and C$_1$ or C$_2$, which is an essential step for HBAC. Fortunately, we show in Section~\ref{sec:ENDOR5q} that it is possible to realize a pulse sequence combined with the electron refresh step in order to polarize both C$_1$ and C$_2$ to the bath polarization, as long as all NMR transitions can be addressed selectively, and the nuclei H and C$_m$ are separately addressable by microwave pulses. \\
\indent The RF pulse durations are typically on the order of tens of $\mu$s, corresponding to excitation bandwidths less than $100$ kHz. Thus, for the 5 qubit sample, we search for an orientation in which all NMR transition frequencies are at least $500$ kHz apart, and all allowed ESR transition frequencies with the exception of the C$_1$ and C$_2$ resonances are at least $12$ MHz apart. Then among this set, we select one orientation in which $\tan(\eta)$ is as small as possible for all nuclei. The optimal orientation for the fully-labeled molecule and relevant parameters in that orientation is given in Table \ref{ENDORorientation}. The minimum distance between ESR transitions in this orientation is $6$ MHz due to C$_1$ and C$_2$. All other ESR transitions are at least $13$ MHz apart from each other. Some forbidden transitions may be very close to the ESR allowed frequencies, but this is not a serious issue since the forbidden transitions can be ignored when $\tan(\eta)$ is small.\\
\indent For the 3 qubit molecule, it is relatively easy to find orientations in which the allowed ESR transitions are separated by at least $12$ MHz. Among that set, an optimal orientation is chosen for which $\tan(\eta)$ is minimized for both H and C$_m$. The results are summarized in Table~\ref{ENDORorientation}. 
\begin{table}[h]
\renewcommand{\arraystretch}{1.2}
\centering
\begin{tabular}{c c|C{1cm}|C{1cm}|C{1cm}|C{1cm}| c}
\cline{3-6}
& &\multicolumn{4}{c|}{$A^n_{zz}$(MHz), $B^n$(MHz), $\tan(\eta^n)$} &\\ \cline{2-7}
\multicolumn{1}{C{1cm}|}{}& $\theta,\phi$ & H & C$_m$ & C$_{1}$&C$_2$& \multicolumn{1}{C{1cm}|}{min$(\vec{\delta})$ (MHz)}\\ \cline{1-7}
\multicolumn{1}{|C{1cm}|}{5 qubit} & $95.8\degree,86.3\degree$ & -56.36, 3.59, 0.048 & 209.48, 22.51, 0.0039 & -43.50, 0.70, 0.0029 & -37.56, 0.70, 0.004 &\multicolumn{1}{C{1cm}|}{6}\\ \hhline{|=|=|=|=|=|=|=|}
\multicolumn{1}{|C{1cm}|}{3 qubit} & $90\degree$, $90\degree$& -56.0, $0$, $0$ & 211.83, 8.98, 0.0015 & & &\multicolumn{1}{C{1cm}|}{56}\\ \cline{1-7}
\end{tabular}
\renewcommand{\arraystretch}{1}
\caption{\label{ENDORorientation} Optimal orientations for the ENDOR control scheme with fully-labeled (5 qubit) and $^{13}$C$_m$-labeled (3 qubit) molecules, corresponding values of hyperfine coupling constants $A^n_{zz}$ and $B^n$, and the quantity characterizing forbidden transition rates $\tan(\eta^n)$ for nuclear spin $n$. In these orientations, the NMR transitions are well separated for high fidelity control; the minimum distances between two NMR transition frequencies are $1.8$ MHz and $7.6$ MHz for the 5 qubit and 3 qubit systems, respectively. Here, min$(\vec{\delta})$ represents the minimum separation between allowed ESR transitions.}
\end{table}

\section{Simulation of Heat-Bath Algorithmic Cooling}
\label{Simulation}
The Partner Pairing Algorithm (PPA) \cite{PPA} is the optimal method for implementing HBAC when there is one reset qubit that thermalizes to the bath polarization $\epsilon_b$ much faster than the relaxation rate of the system qubits. In this scenario, polarizing one system qubit beyond $\epsilon_b$ can be achieved by repeatedly applying the entropy compression and refresh steps. The compression is a permutation that rearranges the diagonal elements of the density matrix in non-increasing order so that the polarization of the $1^{\text{st}}$ (target) qubit increases while the reset qubit polarization decreases. During the refresh step, the reset qubit thermalizes to the bath temperature and the overall entropy of the system is reduced. As this procedure is repeated, the polarization of the target qubit, $\epsilon_t$, asymptotically approaches a threshold value: $\epsilon_{th}=\epsilon_b 2^{n-2}$ if $\epsilon_b\ll 2^{-n}$, and polarization arbitrarily close to unity when $\epsilon_b \gg 2^{-n}$, where $n$ is the number of system qubits used in the compression step~\cite{moussa2005heat,PPA}.\\
\indent We perform simulations of 3-qubit and 5-qubit HBAC for cooling one target qubit below the electronic spin bath temperature using $^{13}$C$_m$-labeled and $^{13}$C$_{m,1,2}$-labeled molecules. The HBAC protocol consists of two steps: (i) apply a swap gate between the electron and each nucleus, with electron refresh steps in between, cooling all nuclei to the bath temperature; (ii) run multi-qubit compression to boost the polarization of the target spin. For the 3 qubit case, the protocol we simulate is equivalent to the PPA. For the 5 qubit simulation, designing the stepwise operations corresponding to the PPA is challenging since each compression gate depends on the input state of the gate and is generally a different gate for each compression step. Moreover, the PPA compression gates may involve complicated pulse sequences. Thus, the 5-qubit protocol simulated here is designed to be analogous to the 3-qubit sequence, i.e. a polarization transfer from the electron to all nuclei followed by a 5-qubit compression step. This same sequence can be repeated until asymptotic polarization is reached, although here we only simulated one round to demonstrate the feasibility of cooling beyond the Shannon bound in this 5-qubit system. \\
\indent Two control methods are tested for the HBAC simulations: AHC and ENDOR. The electronic $T_1$ and $T_2$ processes are modelled as a Markovian dynamical map, and simulated by solving a master equation of the Lindblad form~\cite{dec1,dec2}. Inhomogeneous line broadening of the electron spin resonances, characterized by $T^*_2$, is taken into account by averaging the simulation over a set of spin Hamiltonians in which the magnitude of electron Zeeman energy is a Lorentzian-distributed random variable. We use experimentally measured electron $T_1$ and $T_2$ values to determine the Lindblad operators, and the measured ESR line width to determine the $T^*_2$ Hamiltonian distribution. The electron relaxation parameters are $T_1=27$ $\mu$s, $T_2=5$ $\mu$s, and $T^*_2=28$ ns at room temperature. In the following sections, we describe the simulations in detail and report the results.

\subsection{Simulation of 3-qubit HBAC}
\label{HBAC3q}
The quantum circuit for 3-qubit HBAC is shown in the top part of Fig.~\ref{AHC3qAC}. Here, the electron is chosen as the bath qubit whose polarization is $\epsilon_b\approx 8\times 10^{-4}$ at room temperature and at $B_0 = 2\pi\cdot 10$ GHz$/\gamma_e\approx 3568$ G. Under these conditions, $^{1}$H and $^{13}$C equilibrium polarizations are about 660 and 2620 times smaller than the electronic polarization. For the 3-qubit PPA with completely mixed initial states, the first two entropy compression steps are polarization transfers (swaps) from the reset qubit to the $1^{\text{st}}$ (target) and the $2^{\text{nd}}$ qubits. After the $3^{\text{rd}}$ gate operation (3-qubit compression), the target qubit polarization is higher than the bath polarization. These steps complete one round of the 3-qubit PPA. In theory, the first round of the 3-qubit PPA boosts the polarization of one nuclear spin to $1.5\epsilon_b-0.5\epsilon_b^3\approx1.5\epsilon_b$ for small $\epsilon_b$. By repeated application of these steps, the nuclear spin is asymptotically polarized to $2\epsilon_b$. Note that starting from the $2^{\text{nd}}$ round, only one polarization swap gate is used since the target nucleus is already polarized above the bath polarization.

\subsubsection{Anisotropic hyperfine control}
\label{AHCAC}
\begin{figure}[h]
\centering
\includegraphics[width=0.62\textwidth]{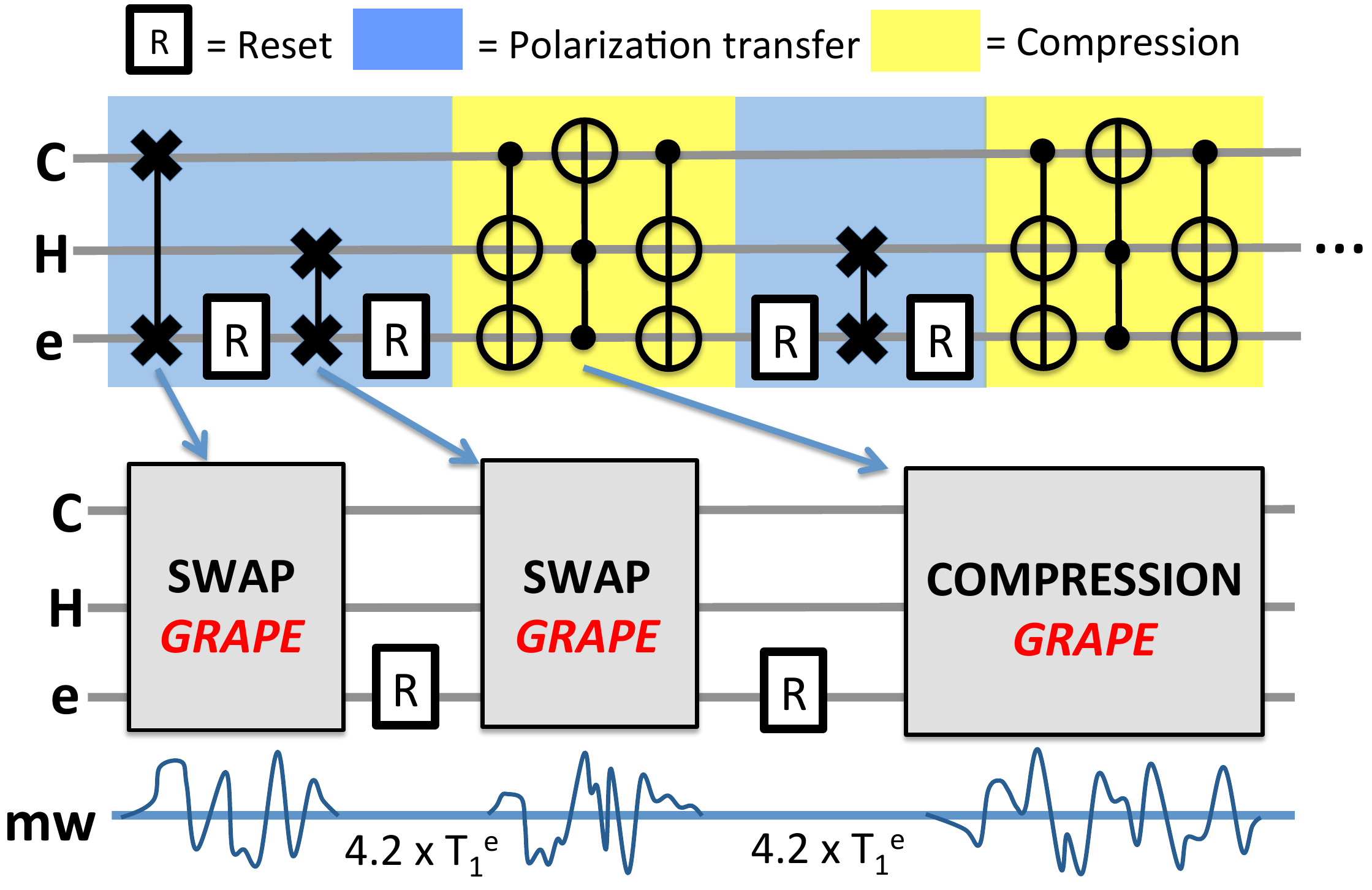}
\caption{\label{AHC3qAC} Quantum circuit of 3-qubit HBAC using the AHC method. The electron spin is the reset qubit, and it is refreshed by waiting for $4.2\times T_1^e$. In the optimal crystal orientation for AHC, the forbidden transition strength of C$_m$ is weaker than that of the proton. Consequently, the C$_m$ polarization decays more slowly during electron reset and C$_m$ is chosen as the target qubit. Two rounds of the algorithm are shown in the top panel. The 3-qubit PPA iteratively applies polarization transfer (shaded in blue) and the compression (shaded in yellow). All gates are designed in the microwave domain using the GRAPE algorithm. The GRAPE pulse lengths are $840$ ns for the swap gates and $900$ ns for the compression gate. A schematic of the AHC sequence for this circuit is shown in the lower part.}
\end{figure}

The crystal orientation used for the 3-qubit experiment using the AHC method is shown in Table~\ref{AHCorientation}. In this orientation, the forbidden transition rate of C$_m$ in the presence of a resonant microwave field is weaker than that of the $\alpha$-proton. Consequently, the C$_m$ polarization decays slower than that of proton during electron reset. Therefore C$_m$ is chosen as the target qubit for cooling. The electron reset is done by waiting for $4.2\times T_1^e$ in order to bring the electron polarization to about $98.5\%$ its thermal polarization. A longer waiting time, e.g. $5\times T_1^e$, brings the electron polarization to above $99\%$, however, we find that the wait time of $4.2\times T_1^e$ is optimal since nuclear spin polarizations decay during electron reset (due to the anisotropic hyperfine interaction). The microwave field swap and compression gates are designed using the gradient ascent pulse engineering (GRAPE) algorithm \cite{khaneja2005optimal}. The GRAPE  pulse lengths are $840$ ns for the swap gates and $900$ ns for the 3-qubit compression gate. The pulses are optimized over the electron Zeeman Hamiltonian distribution in order to be robust to $T^*_2=28$ ns. The design fidelities averaged over this distribution are $99\%$, $98\%$ and $96\%$ for the e-H swap, e-C$_m$ swap, and 3-qubit compression, respectively. The bottom part of Fig.~\ref{AHC3qAC} illustrates the implementation of controls necessary for the first round of 3-qubit HBAC using AHC.

\begin{figure}[h]
\centering
\includegraphics[width=0.55\textwidth]{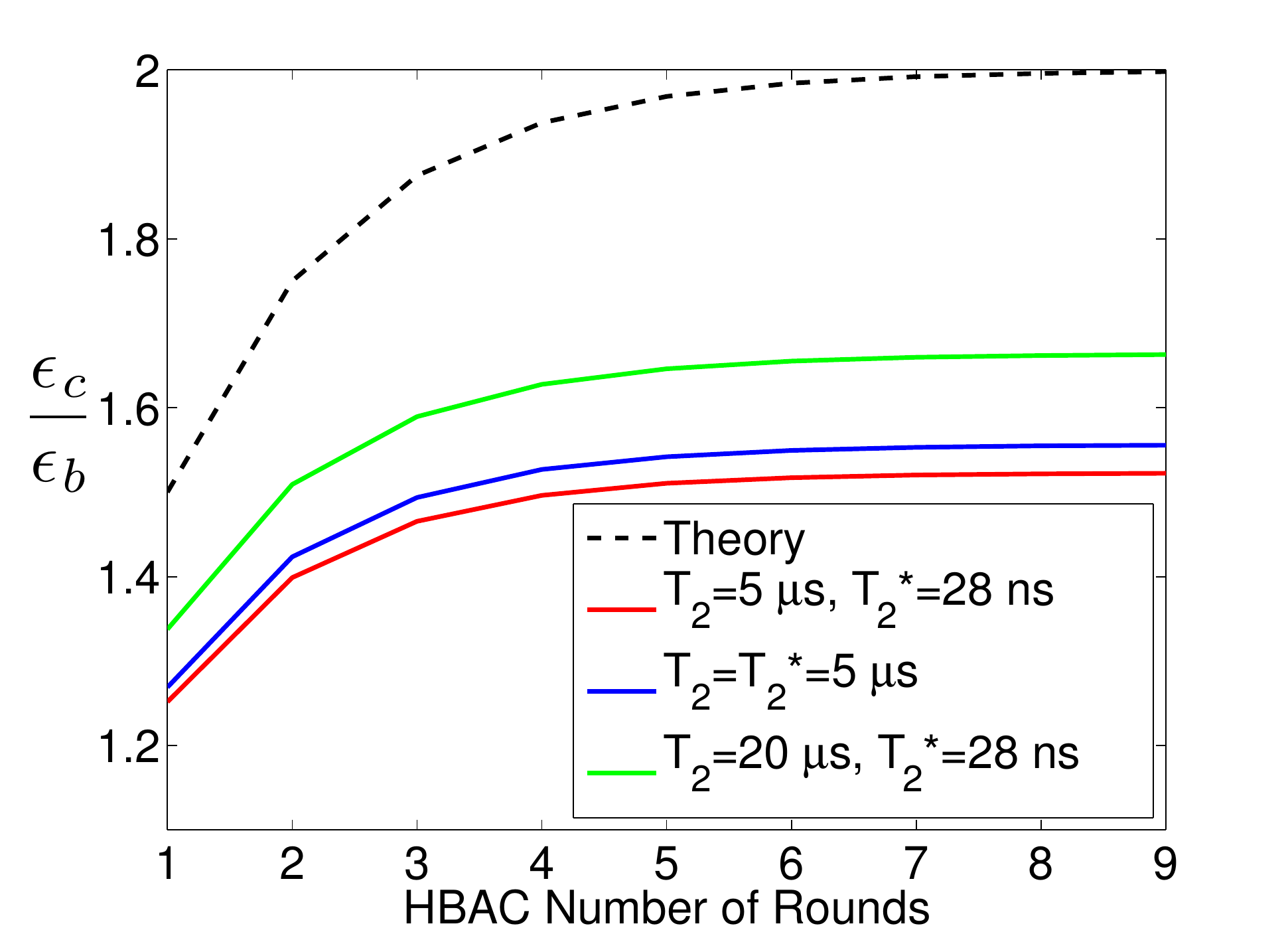}
\caption{\label{AHC3qResult} Simulation results for 3-qubit HBAC using AHC. The plot shows the ratio of the C$_{m}$ polarization ($\epsilon_c$) and the electron bath polarization ($\epsilon_b$) at the end of each HBAC round, up to 9 rounds. The black dashed curve is the theoretical (ideal) value. The red curve is obtained by incorporating all experimentally determined (room temperature) relaxation parameters. The blue curve is obtained by allowing the $T^*_2$ of the electron to equal $T_2^e$. The green curve is obtained by allowing $T_2^e$ to be $20$ $\mu$s, 4 times longer than the measured value, in order to test the consequence of a longer $T_2$.}
\end{figure}

Fig.~\ref{AHC3qResult} shows the polarization of C$_m$ at the end of each round of 3-qubit HBAC. The red curve is obtained when the simulation incorporates the experimentally determined, room temperature electron relaxation effects. The C$_m$ polarization exceeds the bath polarization after the first round of HBAC, and increases further asymptotically as the HBAC is repeated. Nevertheless, the polarization enhancement is below the theoretically calculated value; after 9 HBAC rounds, the polarization of C$_m$ is about $76\%$ of the theoretical value. The largest contribution to the error is the loss of nuclear polarization due to the electron $T_1$ process during the application of GRAPE pulses and the electron refresh steps. In the absence of decoherence, the GRAPE pulses transfer $99.6\%$ and $99.8\%$ of the electron polarization to C$_m$ and H, respectively. The compression gate polarizes the carbon to $1.49\epsilon_b$ at the end of the first round. However, when the $T_1$ of the electron is introduced while $T_2$ and $T^*_2$ are still assumed to be infinite, the carbon and hydrogen polarizations prior to the compression step are reduced to $94.4\%$ and $94.3\%$ of the electron polarization, and the compression yields the carbon polarization of $1.39\epsilon_b$. Another source of the error is the finite ratio of the electron $T_2$ to the pulse duration; $T_2^e$ is only about 5 to 6 times longer than the GRAPE pulses. One can imagine another type of 3-qubit molecule whose electron $T_2$ is longer, for instance, $20$ $\mu$s. Then the polarization of the carbon after 7 rounds of HBAC is $1.66\epsilon_b$, which is $83\%$ of the theoretical value (i.e. the green curve in Fig.~\ref{AHC3qResult}). Finally, we consider a scenario in which the ESR linewidth is much narrower and $T_2$ limited, i.e. $T^*_2=T_2=5$ $\mu$s. This result is indicated in blue in Fig.~\ref{AHC3qResult}. Although there is a slight improvement by having a longer $T^*_2$, the experimental value of $T^*_2$ does not pose a significant problem since the GRAPE pulses are designed to be robust to inhomogeneous line broadening. 

\subsubsection{ENDOR control}
\label{ENDORAC}
\begin{figure}[h]
\centering
\includegraphics[width=0.62\textwidth]{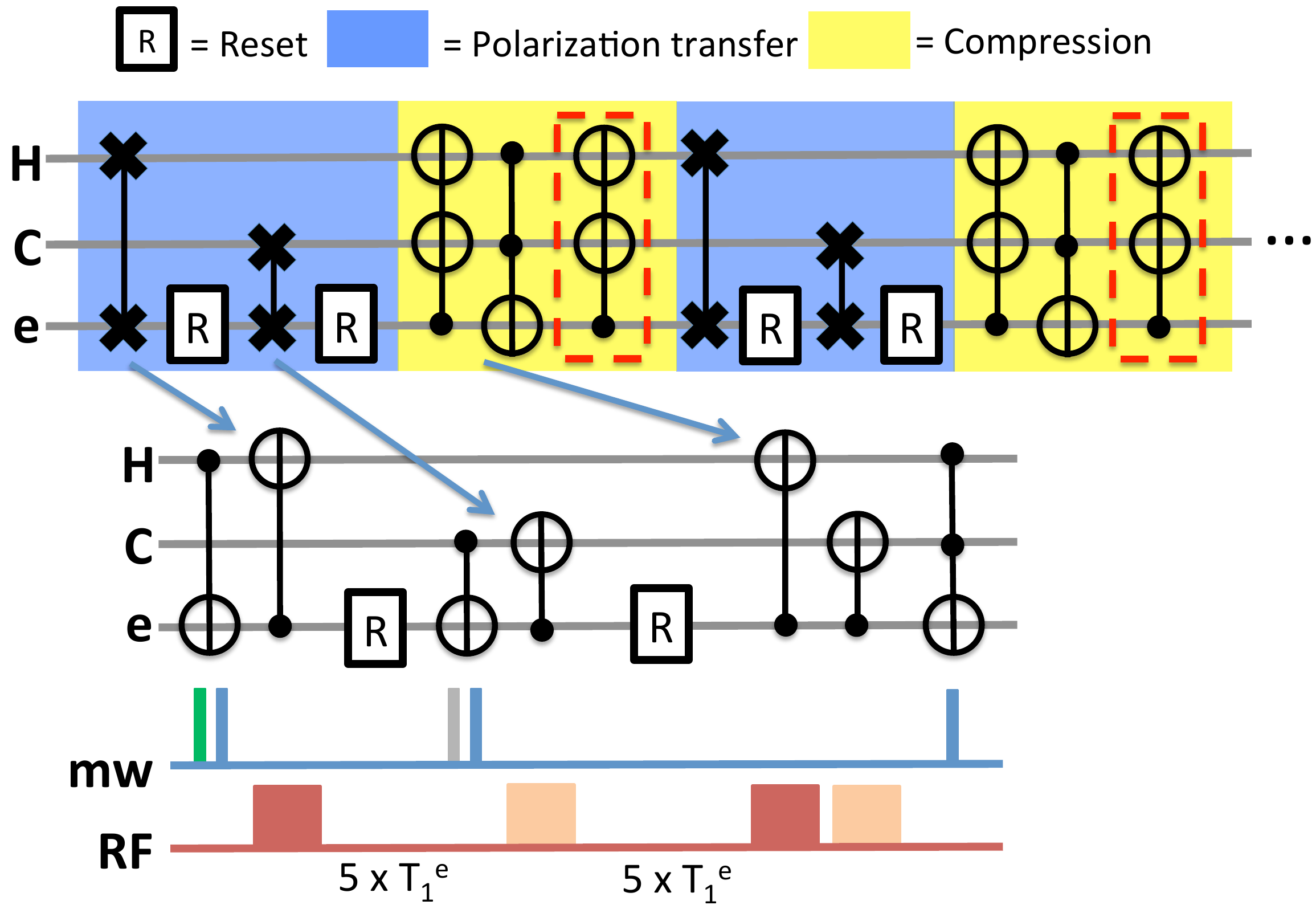}
\caption{\label{ENDOR3qAC} Quantum circuit of the 3-qubit HBAC using ENDOR control. The electron is used as the reset qubit, and it is refreshed by waiting for $5\times T_1$. In the crystal orientation optimized for ENDOR, the forbidden transition rates of H are weaker than those of C$_m$, so H is chosen as the target qubit. The 3-qubit PPA iteratively applies polarization transfer (shaded in blue) and compression (shaded in yellow). All gates are realized using selective microwave and RF $\pi$-pulses. The compression step can only be implemented here to boost the electron polarization (see text for explanation), so a swap gate between the electron and proton is used to store the boosted polarization on the proton. The CNOT gates in the red dashed box are not necessary and are left out of the simulated implementation. The quantum circuit in the middle shows the gate decomposition of the swap and the compression into CNOT and Toffoli gates. The schematic of the microwave and RF pulse sequence at the bottom shows that the CNOT and Toffoli gates can be realized by transition-selective pulses. Shading illustrates different pulse frequencies.}
\end{figure}

In a pulsed ENDOR control scheme, the crystal orientation is chosen as shown in Table~\ref{ENDORorientation}. In this orientation, the forbidden transition rates are both minimized, and it turns out the rate for H is weaker than that of C$_m$. Thus, H is chosen as the target qubit for cooling. The electron reset is done by waiting for $5\times T_1$ in order to bring the electron polarization to $99.3\%$ of the thermal polarization. The loss of nuclear polarization during reset and control operations is negligible since the forbidden transition rates are both very weak. As shown in the bottom part of Fig.~\ref{ENDOR3qAC}, the swap and compression gates can be decomposed into controlled-not (CNOT) gates and a toffoli gate. These operations can be realized by selective microwave and RF $\pi$-pulses. For example, the CNOT gate that flips the H spin if the electron is `spin down' can be realized by RF pulses at frequencies that correspond to $\kket{\uparrow_H\uparrow_C\downarrow_e}\leftrightarrow\kket{\downarrow_H\uparrow_C\downarrow_e}$ and $\kket{\uparrow_H\downarrow_C\downarrow_e}\leftrightarrow\kket{\downarrow_H\downarrow_C\downarrow_e}$ transitions. In the compression step, a Toffoli gate that flips H spin if both C$_m$ and electron are `spin down' is required. However, the Toffoli gate on the proton cannot be realized by this method because the $\kket{\uparrow_H\uparrow_C\downarrow_e}\leftrightarrow\kket{\downarrow_H\uparrow_C\downarrow_e}$ transition frequency is identical to the $\kket{\uparrow_H\downarrow_C\downarrow_e}\leftrightarrow\kket{\downarrow_H\downarrow_C\downarrow_e}$ frequency. On the other hand, a Toffoli gate that flips the electron spin when both H and C$_m$ are `spin down' can be realized by applying a microwave pulse at the frequency of the $\kket{\downarrow_H\downarrow_C\uparrow_e}\leftrightarrow\kket{\downarrow_H\downarrow_C\downarrow_e}$ transition, which is distinct from all other allowed ESR transition frequencies. Therefore, we modify the quantum circuit to extract the entropy from the electron during the compression, and use an additional swap gate to transfer the final polarization to the proton (see Fig.~\ref{ENDOR3qAC}). \\
\indent This quantum circuit enables one to repeatedly apply HBAC for pumping the H polarization. The compression step consists of four CNOT gates and a Toffoli gate, as shown in Fig.~\ref{ENDOR3qAC}.  Since the goal of the compression is only to extract entropy from the electron, the last two CNOT gates targetting the nuclear spins (shown in the red dashed box in Fig.~\ref{ENDOR3qAC}) are not necessary. Hence, the compression is reduced to two CNOT gates and a Toffoli gate as shown in the middle part of Fig.~\ref{ENDOR3qAC}. The rectangular-shaped microwave and RF pulses shown at the bottom of Fig.~\ref{ENDOR3qAC} represent selective $\pi$-pulses, and the different shading of these pulses illustrate different frequencies.

\begin{figure}[h]
\centering
\includegraphics[width=0.52\textwidth]{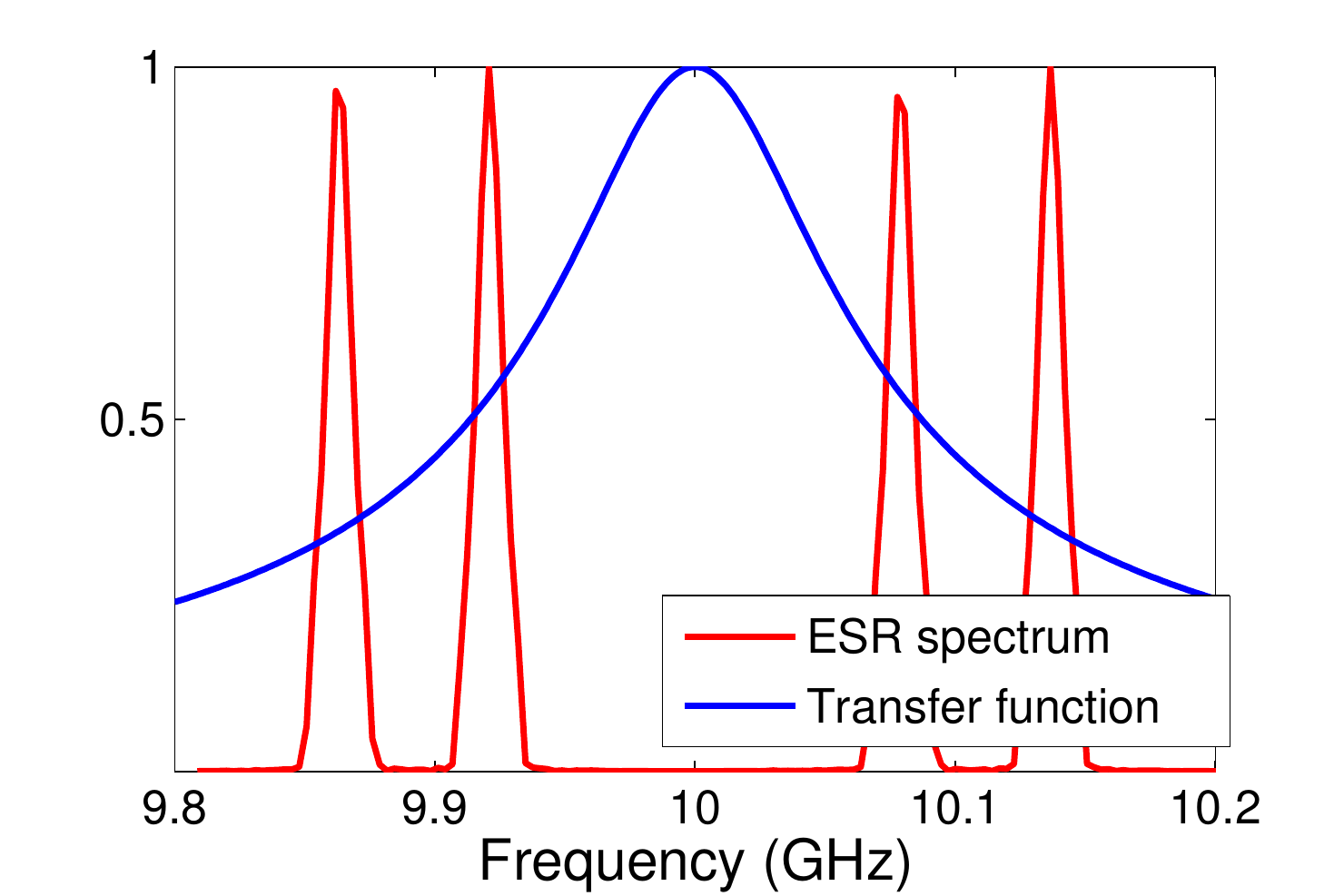}
\caption{\label{transfer} Simulated field-sweep spectra for the $^{13}$C$_m$-labeled malonic acid in the orientation given in Table~\ref{ENDORorientation} (red), and voltage transfer function for a microwave cavity with $Q=100$ (blue). The $y$-axis represents the scaling factor of square root of the microwave power in the cavity a function of the microwave frequency. Given a fixed maximum available microwave power, pulse durations for offset pulses must be increased relative to cavity-resonant pulses.}
\end{figure}

The maximum power for microwave pulses is chosen such that the Rabi frequency is $25$ MHz. In practice, the resonant microwave cavity has a finite bandwidth. This means that more input power is needed in order to drive transitions whose frequency is offset with respect to the cavity resonance frequency. Since the input power is limited in any real experimental setup, the pulse length must increase in order to apply the $\pi$-pulse at the offset frequency. Fig.~\ref{transfer} shows a simulated ESR spectrum centered at the resonator resonance frequency of $10$ GHz (red), and the voltage transfer function for a resonator with quality factor ($Q$) of 100 (blue). If the pulse is applied at a frequency at which the corresponding value of the transfer function is $x$, then the pulse length must increase by a factor of $1/x$ to compensate for the loss in transmitted power. This finite bandwidth effect must be taken into account in simulations because the algorithm cannot be successful if microwave pulse durations become comparable to the electron $T_2$. In the present simulations, the resonator quality factor $Q$ is set to 100 and microwave pulse lengths are adjusted accordingly. The RF $\pi$-pulses are realized using $15$ $\mu$s (H) and $60$ $\mu$s (C$_m$) square pulses, reflecting typical RF amplifier output power levels and assuming an untuned (broadband) RF circuit. 

\begin{figure}[h]
\centering
\includegraphics[width=0.55\textwidth]{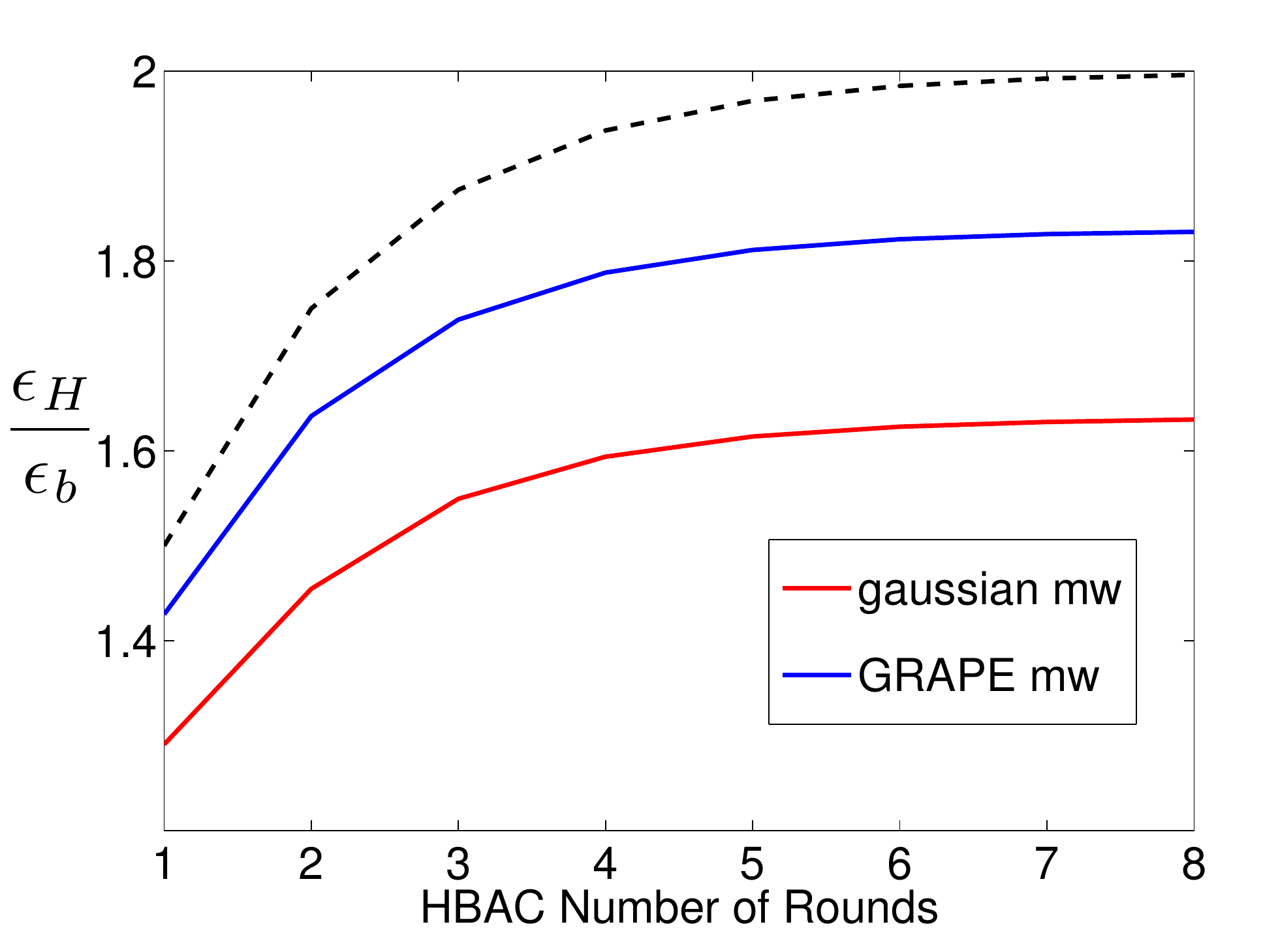}
\caption{\label{ENDOR3qResult} Simulation results of the 3-qubit HBAC using ENDOR control. The plot shows the ratio of H polarization to the electron thermal polarization after each round of the HBAC, up to 8 rounds. The black dashed curve is the theoretical value. The selective microwave $\pi$-pulses are designed in two different ways: (i) $50$ ns gaussian-shaped MW pulses with $20$ ns full width at half maximum (red), (ii) and GRAPE pulses (blue). In both cases, RF transitions are applied using square transition-selective pulses with $15$ ns and $60$ ns pulse lengths for H and C$_m$ spins, respectively.}
\end{figure}

Using the room-temperature electron $T_1$, the HBAC algorithm will not be successful because RF pulse lengths are similar to $T_1$. This can be solved by exploiting longer $T^e_1$ values at lower temperatures. In unlabelled, irradiated malonic acid, we have found experimentally a $T_1$ that grows roughly exponentially with temperature, e.g. with values of 29 $\mu$s, 2.6 ms and 11 ms at room temperature, $43$ K,  and $22$ K, respectively. For simulating 3-qubit HBAC with ENDOR, we choose $T_1 = 2.6$ ms (ignoring the fact that a structural phase transition occurs at 47 K, probably complicating the ESR spectrum). The selective microwave pulses must be designed with a care. While each pulse selectively excites certain transitions, they must also be broad enough to cover the ESR linewidth of about $12$ MHz. First, we simulate $50$ ns gaussian-shaped pulses with the full width at half maximum of $20$ ns in order to excite the entire ESR linewidth and remain selective on the particular transition. The result is shown in Fig.~\ref{ENDOR3qResult} in red. The main source of the error is the inability of the Gaussian pulses to uniformly rotate all spins across the ESR linewidth; the pulse bandwidth must be close to $50$ MHz in order to fully excite the entire ESR line width, but the minimum distance between two ESR allowed transition frequencies is about $56$ MHz. Thus, even in the absence of $T_1$ and $T_2$ effects, each swap gate loses $8\%$ polarization. For the improved microwave control, selective $\pi$-pulses are engineered using the GRAPE algorithm and can be made robust to $T^*_2=28$ ns which corresponds to the ESR line width of $12$ MHz. Using this method, swap gates can transfer $98\%$ of the polarization from the electron to the target nuclear spin. The simulation results for HBAC using GRAPE microwave pulses and rectangular RF pulses are shown in blue in Fig.~\ref{ENDOR3qResult}.

\subsection{Simulating 5-qubit HBAC}
\label{sec:ENDOR5q}
\begin{figure}[h]
\centering
\includegraphics[width=0.62\textwidth]{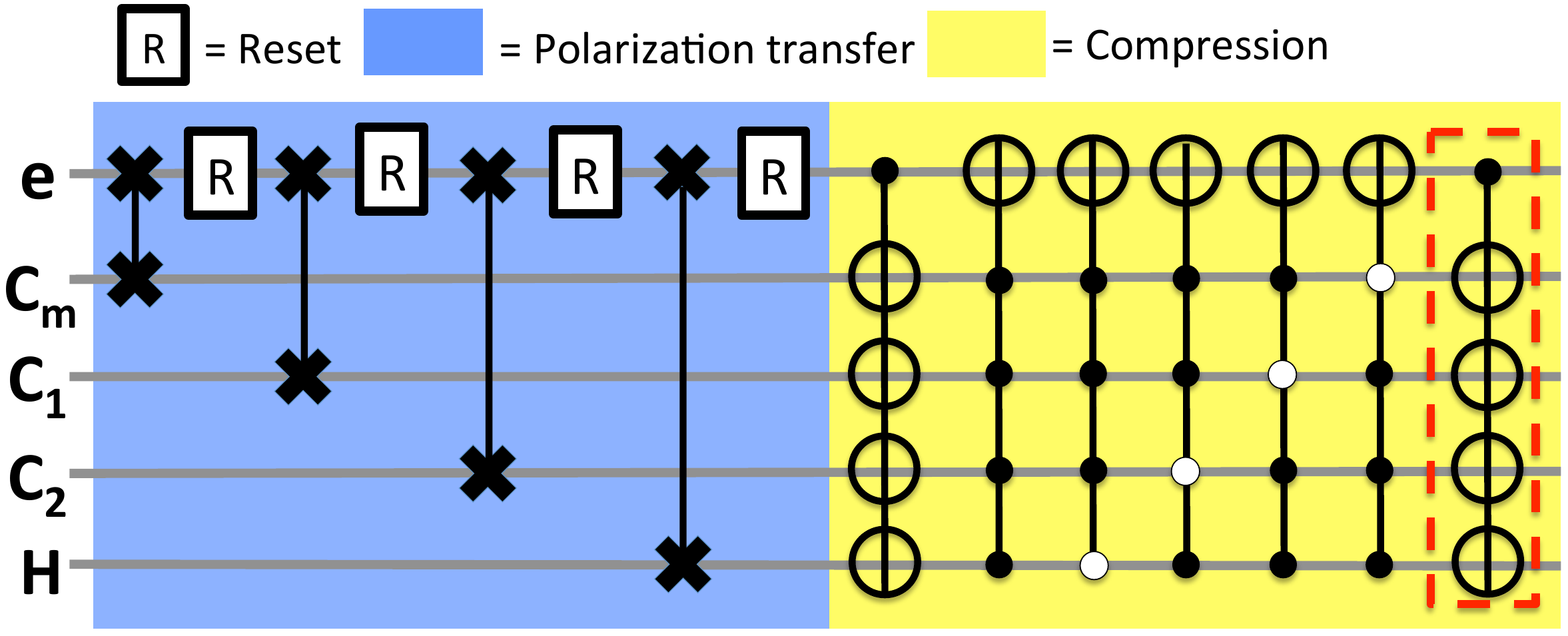}
\caption{\label{ENDOR5qAC} Quantum circuit for 5-qubit HBAC with the electron as the target spin. The polarization can be transferred to a nuclear spin with an additional swap gate. The last CNOT gates inside the red dashed box of the compression step are unnecessary, and are left out of the simulation. The open circles in the controlled gates indicate that the target spin is flipped if the control qubit is in the `spin up' state. One round of the algorithm shown.}
\end{figure}

As discussed in Section~\ref{COAHC}, due to weak forbidden transition rates of C$_{1,2}$ and frequency overlap among ESR transitions, the AHC scheme cannot be implemented in the 5-qubit sample. Therefore, we focus on ENDOR control techniques in the simulation of 5-qubit HBAC. Here, we simulate one round of the 5-qubit HBAC as shown in Fig.~\ref{ENDOR5qAC}, to demonstrate that cooling beyond the Shannon bound is experimentally feasible in the 5-qubit molecule. The quantum circuit we employ for 5-qubit HBAC is shown in Fig.~\ref{ENDOR5qAC}. The electron is the target qubit, but the circuit can be easily modified to cool a nuclear spin by adding a swap gate after the compression step. Theoretically, by applying one round of the quantum circuit, the target qubit is cooled to $\epsilon_b(15-10\epsilon_b^2+3\epsilon_b^4)/8\approx 1.875\epsilon_b$ for small $\epsilon_b$. The orientation of the magnetic field used in the simulation is shown in Table~\ref{ENDORorientation}, and the electron polarization and relaxation parameters are the same as used in the 3-qubit ENDOR HBAC simulation. The forbidden transition rates of the carbons are weaker than those of the proton, so the polarization transfer to H is done just before the compression gate, as shown in Fig.~\ref{ENDOR5qAC}. Similar to the 3-qubit HBAC ENDOR circuit, the last CNOT gates in the compression step (inside the red dashed box) are not necessary since the electron is the target spin. 

\begin{figure}[h]
\centering
\includegraphics[width=0.52\textwidth]{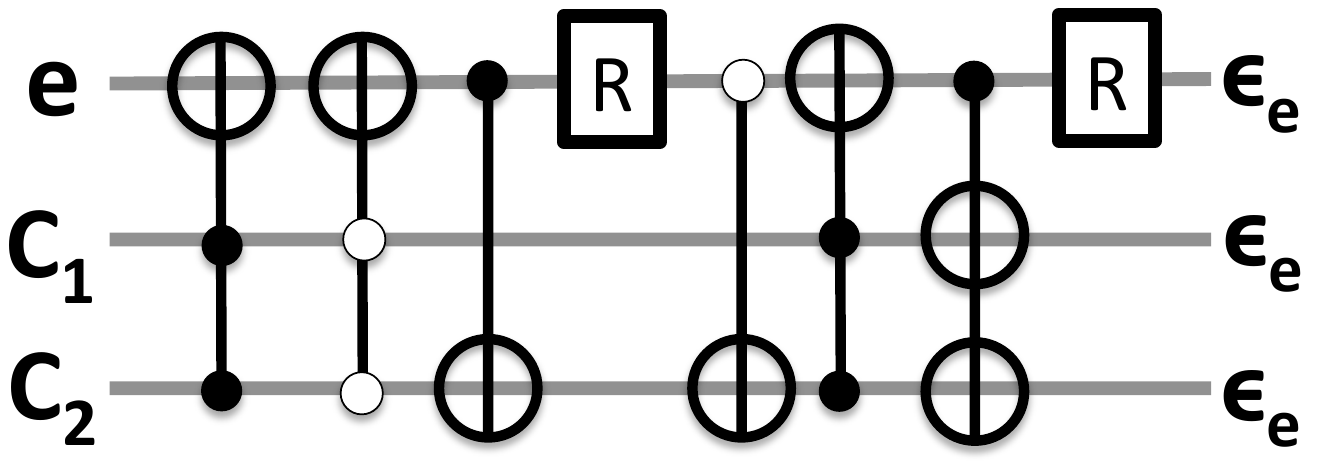}
\caption{\label{eC12swap} Circuit for polarization transfer between the electron and C$_{1,2}$. The circuit brings C$_1$ and C$_2$ to the bath temperature without selective microwave control, but requires selective RF pulses. The open circles in the controlled gates indicate that the target spin is flipped if the control qubit is in the `spin up' state.}
\end{figure}

The usual method for a swap gate between the electron and one of the carboxyl carbons requires selective microwave control of C$_1$ and C$_2$. However, the hyperfine tensors of C$_1$ and C$_2$ are very similar and the two spins cannot be separately addressed by microwave pulses. However, they can be separately addressed with RF pulses (see Section~\ref{OCENDOR}). Therefore, the polarization transfer step for C$_{1,2}$ is modified as shown in Fig.~\ref{eC12swap}, and theoretically polarizes both C$_1$ and C$_2$ to the electron thermal polarization. 

\begin{figure}[h]
\centering
\includegraphics[width=0.55\textwidth]{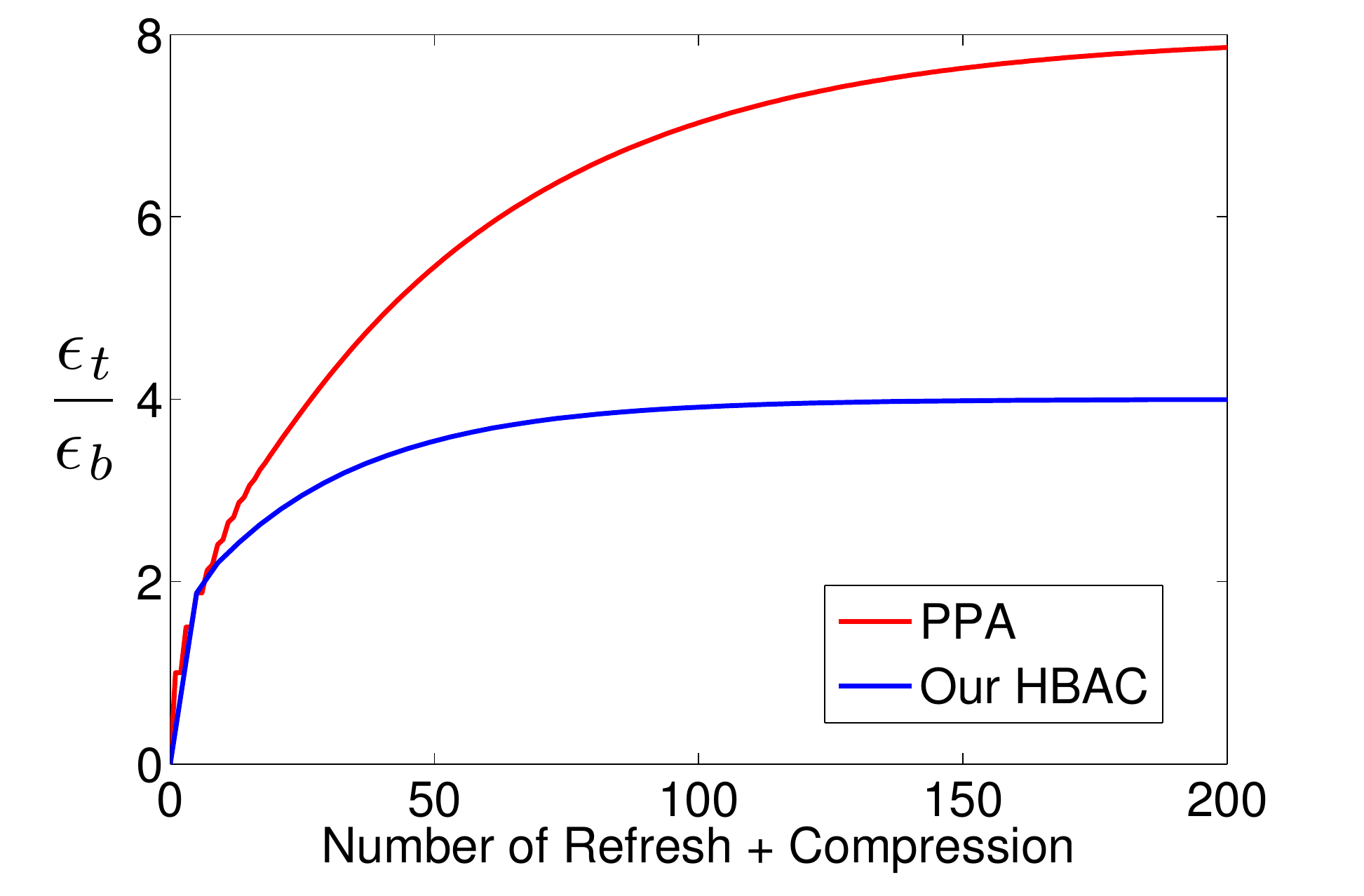}
\caption{\label{PPAvsHBAC5q} Theoretical target qubit polarization $\epsilon_t$ normalized by the bath polarization, as a function of the number of cooling steps, to show the difference between the 5-qubit PPA and the algorithm in Fig.~\ref{ENDOR5qAC}. Each step consists of one refresh operation and one gate on the system qubits. The red curve is obtained by the PPA while the blue curve is obtained by repeated application of the quantum circuit shown in Fig.~\ref{ENDOR5qAC}. In the low polarization regime, the PPA allows the target qubit polarization to asymptotically approach 8 times the bath polarization, and 4 times the bath polarization with our algorithm.}
\end{figure}

Note that the circuit shown in Fig.~\ref{ENDOR5qAC} is not the optimal cooling algorithm, i.e. the PPA. Fig.~\ref{PPAvsHBAC5q} compares the PPA and our cooling algorithm by showing the target qubit polarization $\epsilon_t$ compared to the bath polarization as a function of HBAC steps. Here, each step consists of one refresh operation and one gate on the system qubits (in contrast to the `round' used previously). In the low bath polarization regime, the PPA asymptotically increases the target qubit polarization to 8 times the bath polarization, while repeated application of the algorithm shown in Fig.~\ref{ENDOR5qAC} yields an asymptotic enhancement of 4. While it is possible to find sequences corresponding to the PPA, it is an open question whether such sequences could be practically implemented in this system. For the PPA, the gate decomposition of each system qubit operation depends on the input state, and can result in a complicated pulse sequence. An advantage of the algorithm implemented in our simulation is that the gate decomposition of each step is relatively simple and the asymptotic limit can be reached by simply repeating the quantum circuit.

As seen already in the 3-qubit HBAC simulation, designing microwave pulses that uniformly rotate all spin across the ESR line width while remaining transition-selective is a challenge. For the 5-qubit system, this problem is exacerbated since the ESR transitions are more closely spaced in frequency. In order to obtain the maximum polarization enhancement, GRAPE optimal control must again be used for designing the selective microwave pulses. GRAPE pulses that are robust to the experimentally determined value of $T^*_2$ have been designed with at least 0.95 state fidelity and with $200$ ns pulse lengths for the mw pulses applied to C$_m$ and C$_2$, $900$ ns for C$_1$ and $^{1}$H, and $100$ ns for the 5-qubit compression. RF pulse lengths are chosen as $5$ $\mu$s and $20$ $\mu$s for $^{1}$H and $^{13}$C, respectively. Including the effects of the electron $T_1$, $T_2$, and $T^*_2$, the polarization enhancement after one round of 5-qubit HBAC is $1.67\epsilon_b$. A major contribution to the error is the broad ESR line width compared to the spacing between ESR resonances, which makes realizing high fidelity transition-selective rotations on a time scale short compared to $T_2$ difficult. Despite this, the simulation shows that a single round of 5-qubit HBAC with a realistic control sequence and relaxation yields a polarization enhancement beyond the Shannon bound, and also beyond the capability of a single round of 3-qubit HBAC. Another simulation is performed assuming a much narrower ESR line width, i.e. $T^*_2=T_2$. In this case, all microwave pulses can be designed with $99\%$ unitary fidelity at $100$ ns pulse length, and all nuclei reach at least $97\%$ of the electron polarization during the polarization transfer steps. The polarization improvement after one round of 5-qubit HBAC is $1.79\epsilon_b$. This is $95\%$ of the polarization improvement predicted by the theory, similar to the result of the 3-qubit HBAC simulation using ENDOR control with GRAPE microwave pulses. We expect the single qubit polarization to reach above $3.6\epsilon_b$, $90\%$ of the theoretical value, as the algorithm is repeated, similar to the 3-qubit case. The result shows that for HBAC with many qubits it is crucial to have a sharp ESR linewidth, which motivates a search for electron-nuclear spin systems having narrower ESR lines than the malonyl radical.

The 5-qubit HBAC results are summarized in Table~\ref{5qResult}.

\begin{table}[h]
\renewcommand{\arraystretch}{1.2}
\centering
\begin{tabular}{c| c||c|c|}
\cline{3-4}
\multicolumn{1}{c}{} &\multicolumn{1}{c|}{} &\multicolumn{2}{c|}{\textbf{Simulation}} \\ \cline{2-4}
 & \multicolumn{1}{ c||}{\textbf{Theory}} & $T^*_2 = 5$ $\mu$s & $T^*_2 = 28$ ns\\ \cline{1-4}
\multicolumn{1}{|c |}{$\epsilon_t/\epsilon_b$} & 1.875 & 1.79 & 1.67 \\ \cline{1-4}
\end{tabular}
\renewcommand{\arraystretch}{1}
\caption{\label{5qResult} Polarization of the target qubit compared to the bath qubit after one round of 5-qubit HBAC. The quantum circuit for one round of 5-qubit HBAC is shown in Fig.~\ref{ENDOR5qAC}. Results are shown for two different values of $T^*_2$. The electron spin lattice relaxation time is $T_1 = 2.6$ ms in this simulation.}
\end{table} 

Finally, we note two experimental caveats that are not taken into account in our simulations, but do not invalidate the main results. In section~\ref{Simulation}, we accounted for a type of cross-relaxation in which the nuclear polarization decays due to a combination of the electronic spin-lattice relaxation process and the anisotropic term in the spin Hamiltonian of the form $\hat{S}_z\hat{I}_x$. There is an additional cross-relaxation mechanism involving noise acting directly on the nuclear operators, which we have not included in the Lindblad master equation. In future work, we plan to experimentally measure the cross-relaxation rates as a function of orientation and temperature to determine these additional contributions. If the additional mechanism is dominant, going to lower temperatures may be required in order to achieve sufficiently long nuclear $T_1$ timescales. Secondly, a well known structural phase transition takes place in malonic acid at 46 K \cite{krzystek1995ednmr,fukai1991thermodynamic,mccalley1993endor}, which we have not considered here. Below 46 K, the ${\rm P}\bar1$ crystal symmetry is broken and there are two magnetically distinguishable molecules per unit cell. We have observed certain ENDOR transitions to split into two below 46 K, consistent with this phase change. However, experiments may be designed using phase-cycling techniques in order to cancel signal contributions from one of the molecules in the unit cell, so this does not in principle prevent the proposed AC experiments from being carried out. 

%
%
\section{Conclusion}
\label{Conclusion}
\indent HBAC is an open-system cooling method that allows at least one qubit to be polarized beyond the Shannon bound, starting from a set of initially mixed qubits and exploiting a controlled interaction with a heat-bath. This is a promising tool for dynamically preparing ancilla qubits to a sufficient level of purity for quantum error correction. In this work, we accurately determined the full spin Hamiltonian of a five qubit electron-nuclear hyperfine coupled system in single crystal, irradiated malonic acid. Using the hyperfine tensors, we determined the optimal magnetic field orientation for achieving high fidelity control using two control methods: anisotropic hyperfine control (AHC) and pulsed ENDOR techniques. Computer simulations were carried out using realistic experimental conditions and including the relevant electron spin relaxation parameters, demonstrating that the realization of 3-qubit and 5-qubit HBAC is feasible in this system. Using the 3-qubit molecule with AHC, the polarization of a nuclear spin is predicted to increase above $1.5\epsilon_b$ after 9 rounds of the cooling algorithm. The ENDOR simulation assumed a lower temperature in which the electron $T_1=2.6$ ms, in order to have the $T_1$ sufficiently long compared to RF pulse lengths. Using GRAPE for selective microwave pulse design, the polarization of a nuclear spin is predicted to increase above $1.8\epsilon_b$ after 9 rounds of the cooling algorithm. There is a tradeoff here between experimental simplicity and the achievable fidelity; the AHC experimental setup is simpler and can be performed at room temperature, but in general yields lower fidelities than the ENDOR approach at low temperature. \\
\indent The carboxyl carbons in the 5-qubit sample have very weak forbidden transition rates under an applied microwave field, which prevents them from being practically controllable using AHC. Hence, for the 5-qubit HBAC simulation, we focused on the ENDOR control scheme. The 5-qubit simulation was again carried out at low temperature for a longer $T_1$, and the selective microwave pulses were designed using the GRAPE method. After one round of 5-qubit HBAC, the simulation predicts that the target qubit (the electron) would reach a polarization of $1.67\epsilon_b$. The major obstacle to reaching the ideal value of $1.875\epsilon_b$ is the small spacing between certain allowed ESR transitions, similar in order to the ESR line width. Another simulation was performed assuming $T^{*}_2=T_2$ to predict the outcome of the algorithm given a molecule that is similar to the malonic radical but with much sharper ESR transitions. Here, a target qubit polarization of $1.79\epsilon_b$ is obtained after one round of the algorithm. \\
\indent We conclude that the experimental demonstration of 3- and 5-qubit algorithmic cooling beyond the Shannon bound is feasible in the isotopically labelled malonyl radical. The 5-qubit system can yield a larger polarization enhancement compared to the 3-qubit system, as expected. The experimental value of $T^{*}_2$ is found to be a critical factor that limits the fidelity of gate operations, and therefore the achievable polarization under HBAC. Nearly ideal results are obtainable when $T^{*}_2=T_2$. 

%
%
\begin{acknowledgements}
This research was supported by NSERC, the Canada Foundation for Innovation, CIFAR, the province of Ontario and Industry Canada. T. Takui would like to thank the support via Grants-in-Aid for Scientific Research on Innovative Areas ``Quantum Cybernetics" and Scientific Research (B) from MEXT, Japan. The support for the present work by the FIRST project on ``Quantum Information Processing" from JSPS, Japan and by the AOARD project on ``Quantum Properties of Molecular Nanomagnets" (Award No. FA2386-13-1-4030) is also acknowledged. We acknowledge David Cory and Richard Oakley for access to CW ESR spectrometers, Aaron Mailman for help with CW ESR, Jalil Assoud for assistance with x-ray spectroscopy, and Robert Pasuta for help with gamma-irradiation of samples. We thank Dawei Lu, Tal Mor and Yossi Weinstein for helpful discussions. 
\end{acknowledgements}

%
%
\bibliographystyle{spphys}

\end{document}